\documentclass[%
 reprint,
superscriptaddress,
 amsmath,amssymb,
 aps,
 pra,
 showkeys,
 floatfix,
]{revtex4-2}

\usepackage{multirow} 
\usepackage[utf8]{inputenc}
\usepackage[T1]{fontenc}
\usepackage{graphicx}
\usepackage{dcolumn}
\usepackage{bm}
\usepackage{booktabs}
\usepackage{array}
\usepackage[colorlinks=true, linkcolor=blue, citecolor=blue, urlcolor=blue]{hyperref}
\usepackage[capitalise]{cleveref}
\usepackage[all]{hypcap}
\usepackage{booktabs} 
\usepackage{amsmath}
\usepackage{amssymb}
\usepackage{mathtools}
\usepackage{mathrsfs}
\usepackage{amsthm}
\usepackage{amssymb}
\usepackage{url}
\usepackage{makecell}
\usepackage{xcolor,soul}
\sethlcolor{yellow}
\usepackage[ruled,vlined,linesnumbered]{algorithm2e}
\soulregister\cite7
\usepackage{float}

\begin{document}

\preprint{APS/123-QED}

\title{All-valid-state HOBO encoding for constrained combinatorial optimization on NISQ devices}

\author{Juncheng Wang}
\affiliation{Department of Electrical Engineering and Computer Science,
Tokyo University of Agriculture and Technology, Koganei, Tokyo 184-8588, Japan}

\author{Takumi Kanezashi}
\affiliation{Department of Electrical Engineering and Computer Science,
Tokyo University of Agriculture and Technology, Koganei, Tokyo 184-8588, Japan}

\author{Daisuke Tsukayama}
\affiliation{Department of Electrical Engineering and Computer Science,
Tokyo University of Agriculture and Technology, Koganei, Tokyo 184-8588, Japan}

\author{Koki Awaya}
\affiliation{Department of Electrical Engineering and Computer Science,
Tokyo University of Agriculture and Technology, Koganei, Tokyo 184-8588, Japan}

\author{Reo Saito}
\affiliation{Department of Electrical Engineering and Computer Science,
Tokyo University of Agriculture and Technology, Koganei, Tokyo 184-8588, Japan}

\author{Jun-ichi Shirakashi}
\email[Contact author: ]{shrakash@cc.tuat.ac.jp}
\affiliation{Department of Electrical Engineering and Computer Science,
Tokyo University of Agriculture and Technology, Koganei, Tokyo 184-8588, Japan}

\author{Tetsuo Shibuya}
\affiliation{Division of Medical Data Informatics, Human Genome Center,
The Institute of Medical Science, The University of Tokyo,
Minato, Tokyo 108-8639, Japan}

\author{Hiroshi Imai}
\affiliation{Graduate School of Information Science and Technology,
The University of Tokyo, Bunkyo, Tokyo 113-8656, Japan}

\date{\today}

\begin{abstract}
Continued advancements in quantum computing have stimulated growing interest in translating quantum technologies into real-world applications. Consequently, the investigation of practically motivated NP-hard problems is of significant value. This study investigates the performance of a variational quantum eigensolver (VQE) in addressing the traveling salesperson problem (TSP) through noiseless simulations representative of noisy intermediate-scale quantum (NISQ) devices using higher-order binary optimization (HOBO) encodings. We construct a HOBO Hamiltonian with an efficient binary representation and propose an all-valid-state HOBO (AVS-HOBO) scheme based on cyclic mapping that eliminates one penalty term and reuses states that would otherwise be invalid. Using TSP instances of up to 20 cities, we compare the original HOBO and AVS-HOBO encodings from multiple perspectives, including the energy convergence behavior and the approximation, tour-length, and feasibility ratios. In addition to simulations, we perform computations on real quantum hardware with different device architectures, where we not only compare the performances of different chips but also investigate the effects of different error-mitigation methods on actual quantum machines. The results indicate that AVS-HOBO encoding enhances the practical reliability of VQE on NISQ devices and improves scalability for larger TSP instances, with broader applicability to constrained quantum optimization problems.
\end{abstract}

\keywords{Quantum Algorithm, NISQ Device, Traveling Salesperson Problem, Variational Quantum Eigensolver, Quantum Processor}

\maketitle
\raggedbottom

\section{Introduction}\label{sec1}

As quantum computer technology advances, the application of quantum computation to real-world problems has become a priority. However, current quantum devices have very limited resources, making algorithm and hardware selection critical. Variational quantum algorithms (VQAs) \cite{cerezo2021variational} stand out as they do not require deep and highly complex quantum circuits and are suitable for both numerical simulations and implementation on noisy intermediate-scale quantum (NISQ) devices \cite{preskill2018quantum,bharti2022noisy,moll2018quantum}. Thus, we aim to realize practical quantum applications based on NISQ devices. As a representative real-world optimization task, we select the traveling salesperson problem (TSP), a classical NP-hard problem. The TSP can be extended to practical applications, including logistics, transportation, aviation, and port operations. For classical computers, the solution space is extremely large, and the computational complexity is high. Quantum computation can potentially explore such huge solution spaces more efficiently, making it a promising candidate for solving combinatorial optimization problems \cite{perez2024variational,nannicini2019performance}.

Previous studies have proposed quadratic unconstrained binary optimization (QUBO) \cite{glover2022quantum,lucas2014ising,salehi2022unconstrained,palackal2023quantum} and higher-order binary optimization (HOBO) \cite{domino2022quadratic,chai2023optimal} encodings for the TSP \cite{glos2022space,schnaus2024efficient}. Compared with QUBO encoding, HOBO encoding can significantly reduce the number of required qubits. Previous work \cite{glos2022space} investigated the quantum approximate optimization algorithm (QAOA) \cite{farhi2014quantum,farhi2017quantum,qian2023comparative} as a representative VQA. In addition, the variational quantum eigensolver (VQE) is another well-known variational framework \cite{peruzzo2014variational,tilly2022variational,mcclean2016theory,tsukayama2025enhancing}. Notably, QAOA can also be regarded as a problem-inspired ansatz within the broader VQE framework for combinatorial optimization. In this broader variational framework, the objective is to minimize the expectation value of the problem Hamiltonian. A recent study on solving the TSP using HOBO encoding indicated that, for the considered settings, a more general VQE approach can perform better than the QAOA-based approach \cite{schnaus2024efficient}. However, for both QUBO and HOBO encoding, selecting the appropriate strengths for the constraint terms remains challenging. This study investigates TSP using the variational quantum eigensolver (VQE) and proposes a new all-valid-state HOBO (AVS-HOBO) encoding, systematically comparing it with the original HOBO encoding. Building upon the original HOBO formulation, we introduce a cyclic compilation scheme that eliminates one TSP constraint and reduces the number of infeasible solutions in the Hilbert space. Unlike previous studies limited to small problem sizes \cite{schnaus2024efficient}, we analyze instances with up to 20 cities. The AVS-HOBO encoding demonstrates improved stability and converges more reliably to feasible solutions across various instances as the problem size increases.

The remainder of this paper is organized as follows. First, we introduce the TSP, detailing its encoding using QUBO, HOBO, and AVS-HOBO formulations. Next, we describe the employed VQE algorithm. We then present numerical simulation results and experimental results obtained on real quantum hardware, and we analyze the performances of the different encodings. Finally, we summarize our findings and discuss directions for future research.

\section{Methodology}\label{sec2}

This section first introduces the TSP model considered in this study and then discusses three encoding schemes for mapping the problem to a Hamiltonian. The first scheme, referred to as QUBO encoding, treats each city as a one-hot vector such that the route over the entire map can be mapped to a quantum state. The second scheme is HOBO encoding, which abandons the one-to-one correspondence property of the QUBO representation and instead groups qubits such that the state of each group forms a binary integer corresponding to a city. This is called binary encoding. The third scheme builds upon the HOBO encoding and optimizes its constraint terms. The introduction of a cyclic compilation scheme ensures that all cities represented by quantum states are valid. We refer to this method as AVS-HOBO encoding.

\subsection{Traveling salesperson problem}\label{subsec2.1}

For a particular set of cities, the TSP seeks a closed route that starts and ends in the same city, where each city is visited exactly once and the total route length is minimized. Consider a TSP instance with $N$ cities, where the total cost of the route can be written as
\begin{equation}
D_{\mathrm{TSP}}=\min \sum_{i=1}^{N}\sum_{\substack{j=1 \\ j\neq i}}^{N} W_{ij}x_{i,j},
\label{eq1}
\end{equation}

\noindent
where $x_{i,j}\in[0,1]$ is a binary variable that equals 1 if the edge from city $i$ to city $j$ is included in the route and 0 otherwise. The distance between cities $i$ and $j$ is denoted as $W_{ij}$, where we assume $W_{ij}=W_{ji}$ and $W_{ii}=0$.

\subsection{QUBO encoding for the TSP}\label{subsec2.2}

\begin{figure}[!t]
\centering
\includegraphics[width=1\columnwidth]{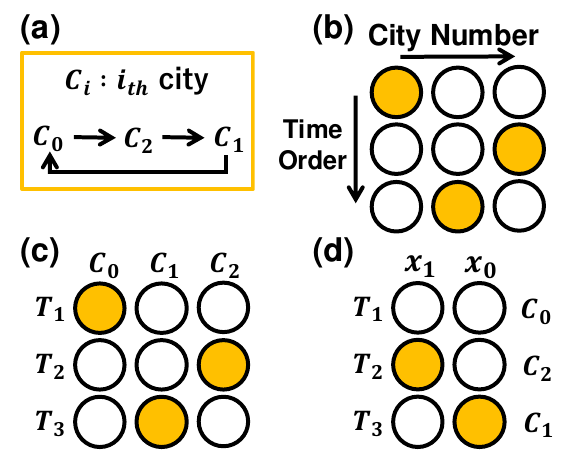}
\caption{Hamiltonian construction method. (a) One possible route for a 3-city TSP. (b) Mapping rule from the TSP to qubits. (c) QUBO encoding and (d) HOBO encoding for the tour in (a). In the QUBO encoding, each qubit corresponds to a city index, whereas in the HOBO encoding, the binary string formed by all qubits at each time step represents a city index. Filled and open circles denote 1 and 0, respectively.}\label{fig1}
\end{figure}

Fig.~\ref{fig1}(a) and (b) illustrate the basic rules of the encoding method using a 3-city TSP example. For the TSP, a widely used approach is to formulate the problem as a QUBO \cite{glos2022space} (Fig.~\ref{fig1}(c)). In this formulation, the visiting order and city index are mapped onto a Hamiltonian in a grid-like manner. For each time step $t\in\{1,\ldots,N-1\}$, we introduce a binary variable $x_{i,t}$ that encodes the visiting order and the city: $x_{i,t}=1$ if city $i$ is visited at time step $t$, and $x_{i,t}=0$ otherwise. QUBO encoding can be written as

{\small
\begin{equation}
\begin{aligned}
H^{\mathrm{QUBO}}(x)=&
\,P_{1}\sum_{t=0}^{N-1}
\left(1-\sum_{i=0}^{N-1}x_{i,t}\right)^{2} \\
&+P_{2}\sum_{i=0}^{N-1}
\left(1-\sum_{t=0}^{N-1}x_{i,t}\right)^{2} \\
&+\sum_{\substack{i,j=0 \\ i\neq j}}^{N-1}
W_{ij}\sum_{t=0}^{N-1}x_{i,t}x_{j,t+1}.
\end{aligned}
\label{eq2}
\end{equation}
}

\noindent
where $P_{1}$ and $P_{2}$ are penalty coefficients set to the same value, and $W_{ij}$ denotes the distance between cities $i$ and $j$. The first term ensures that exactly one city is visited at each time step, whereas the second term ensures that each city is visited exactly once during the entire route. As the distances between cities differ from instance to instance, we choose $P_{1}=P_{2}$ as the maximum distance between any pair of cities in a certain TSP instance. For convenience, we label both cities and the time steps starting from 0. Consequently, this QUBO formulation requires $N$ city indices over $N$ time steps, yielding $N^{2}$ binary variables and thus $N^{2}$ qubits.

In this encoding, the Hilbert space spanned by these $N^{2}$ qubits contains $2^{N^{2}}$ basis states, whereas the number of valid TSP routes is only $N!$. Therefore, the fraction of feasible states in the Hilbert space is
\begin{equation}
R_{\mathrm{fea}}^{\mathrm{QUBO}}=\frac{N!}{2^{N^{2}}}.
\label{eq3}
\end{equation}

\noindent
As the number of cities increases, the proportion of feasible solutions in the Hilbert space decreases exponentially.

\subsection{HOBO encoding for the TSP}\label{subsec2.3}

In the QUBO encoding, each city is represented by a one-hot vector. As an alternative, HOBO encoding has been proposed, where each city is represented as a binary number via a binary encoding scheme \cite{glos2022space}. An example is illustrated in Fig.~\ref{fig1}(d). This encoding further reduces the size of the Hilbert space and thus enables a more efficient representation. HOBO encoding can be written as
\begin{equation}
\small
\begin{aligned}
H^{\mathrm{HOBO}}(b)=&
\,A_{1}\sum_{t=0}^{N-1}
H_{\mathrm{valid}}(b_{t}) \\
&+A_{2}\sum_{0\leq t<t'\leq N-1}
H_{\mathrm{different}}(b_{t},b_{t'}) \\
&+\sum_{i,j=0}^{N-1}
W_{ij}\sum_{t=0}^{N-1}
H_{\delta}(b_{t},i)H_{\delta}(b_{t+1},j).
\end{aligned}
\label{eq4}
\end{equation}

\noindent
where $t$ and $t'$ denote the time steps in the TSP route, and $i$ and $j$ denote the city indices. The binary variable $b_t$ represents the city encoded at position $t$ in the route, whereas $b_{t+1}$ represents the city visited at the next position. The coefficient $W_{ij}$ denotes the travel cost from city $i$ to city $j$. In addition, $A_1$ and $A_2$ are penalty coefficients for the validity and non-repetition constraints, respectively. At each time step, binary string $b_{t}$ is interpreted as an integer that points to a single city. Consequently, the number of qubits required to represent the city index at each time step is reduced to $K=\lceil \log_{2} N \rceil$. Therefore, the possibility of visiting multiple cities simultaneously need not be considered. However, for a fixed-length binary representation, certain bit strings may correspond to invalid city indices. To suppress such invalid encodings, $H_{\mathrm{valid}}^{\mathrm{HOBO}}$ is introduced as a penalty term.
{\small
\begin{equation}
H_{\mathrm{valid}}^{\mathrm{HOBO}}(b_{t})
\coloneqq
\sum_{k_{0}\in K_{0}} b_{t,k_{0}}
\prod_{k=k_{0}+1}^{K-1}
\left[1-\left(b_{t,k}-\tilde{b}_{k}\right)^{2}\right],
\label{eq5}
\end{equation}
}

\noindent
where $\tilde{b}_{K-1}\cdots \tilde{b}_{0}$ is the binary string representing the maximum valid city index. By comparing the binary string at each time step with this bound, we penalize city indices that fall outside the valid range. Another penalty term, $H_{\mathrm{different}}^{\mathrm{HOBO}}$, is used to avoid visiting the same city at different time intervals. Its Hamiltonian is denoted by
\begin{equation}
H_{\mathrm{different}}^{\mathrm{HOBO}}(b,b^{\prime})\coloneqq \prod_{k=0}^{K-1}\left[1-\left(b_{k}-b_{k}^{\prime}\right)^{2}\right].
\label{eq6}
\end{equation}

\noindent
From HOBO encoding, we observe that this approach significantly reduces the number of required qubits; only $N\times\lceil \log_{2} N\rceil$ qubits are required. Correspondingly, the binary encoding reduces the dimensions of the Hilbert space to $2^{N\cdot \lceil \log_{2} N\rceil}$, and the fraction of feasible states in the Hilbert space becomes
\begin{equation}
R_{\mathrm{fea}}^{\mathrm{HOBO}}=\frac{N!}{2^{N\cdot \lceil \log_{2} N\rceil}}.
\label{eq7}
\end{equation}

\noindent
Thus, HOBO encoding reduces the qubit resources while increasing the probability of finding feasible solutions.

\subsection{All-valid-state HOBO encoding for the TSP}\label{subsec2.4}

Although HOBO encoding already reduces the resource requirements and shrinks the Hilbert space compared with QUBO encoding, the constraint term $H_{\mathrm{valid}}^{\mathrm{HOBO}}$ can be further optimized by modifying the encoding scheme. We propose an AVS-HOBO encoding method based on cyclic mapping.

\begin{figure}[!t]
\centering
\includegraphics[width=1\columnwidth]{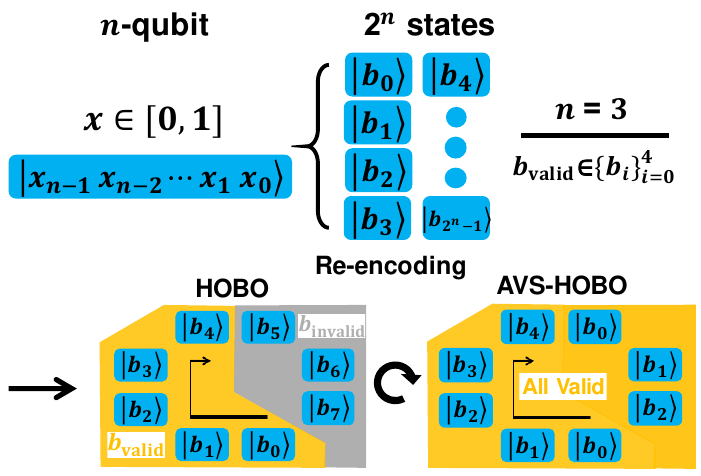}
\caption{Cyclic mapping used in AVS-HOBO to replace the constraint term $H_{\mathrm{valid}}^{\mathrm{HOBO}}$. Here, $n$ is the number of qubits; $n=3$ can generate 8 distinct states. $b_{i}$ denotes the city index. In this figure, we illustrate the case of 5 cities. $b_{\mathrm{valid}}$ denotes valid states, whereas $b_{\mathrm{invalid}}$ denotes unusable states.}\label{fig2}
\end{figure}

We consider a TSP instance with $N$ cities. The length of the binary string required to represent a city index is $K=\lceil \log_{2} N\rceil$, each city index is encoded by $K$ qubits, and there are $M=2^{K}$ possible binary strings in total. When $M>N$, the labels are divided into two parts: the valid region $b_{\mathrm{valid}}\in[b_{0},\ldots,b_{N-1}]$, which corresponds to the required city indices, and the invalid region $b_{\mathrm{invalid}}\in[b_{N},\ldots,b_{M-1}]$. For an invalid label $b_{\mathrm{invalid}}$, a cyclic mapping defined by modulo $N$ is applied:
\begin{equation}
\begin{aligned}
b_{\mathrm{invalid}}^{\prime}
&=b_{\mathrm{invalid}}\bmod N \\
&\in[b_{0},\ldots,b_{m}],
\qquad m=(M-1)\bmod N .
\end{aligned}
\label{eq8}
\end{equation}

\noindent
thus, even if the measurement outcome corresponds to an originally invalid bit string, it is cyclically mapped back to a valid city index, and no penalty is imposed. This enables more efficient reuse of the invalid states. In Fig.~\ref{fig2}, we present examples for a 5-city TSP. In this case, $N=5$, and under the binary representation, $\lceil \log_{2} N\rceil=3$ qubits are required to represent each city. With this cyclic mapping, the constraint term with coefficient $A_{1}$ is no longer required, and the AVS-HOBO Hamiltonian becomes

\begin{equation}
\begin{multlined}
H^{\mathrm{AVS\text{-}HOBO}}(b)=
\,A\sum_{t=0}^{N-1}
\sum_{t^{\prime}=t+1}^{N-1}
H_{\mathrm{different}}(b_{t},b_{t^{\prime}}) \\
+\sum_{i=0}^{N-1}\sum_{j=0}^{N-1}
W_{ij}\sum_{t=0}^{N-1}
H_{\delta}(b_{t},i)H_{\delta}(b_{t+1},j).
\end{multlined}
\label{eq9}
\end{equation}

\noindent
By removing one constraint term, we obtain a new effective solution space. Although the number of qubits required is the same as that in HOBO encoding, the number of possible solutions is effectively reduced to $N^{N}$. Consequently, the fraction of feasible states in AVS-HOBO encoding is
\begin{equation}
R_{\mathrm{fea}}^{\mathrm{AVS\text{-}HOBO}}=\frac{N!}{N^{N}}.
\label{eq10}
\end{equation}

\begin{figure*}[!t]
\centering
\includegraphics[width=1\textwidth]{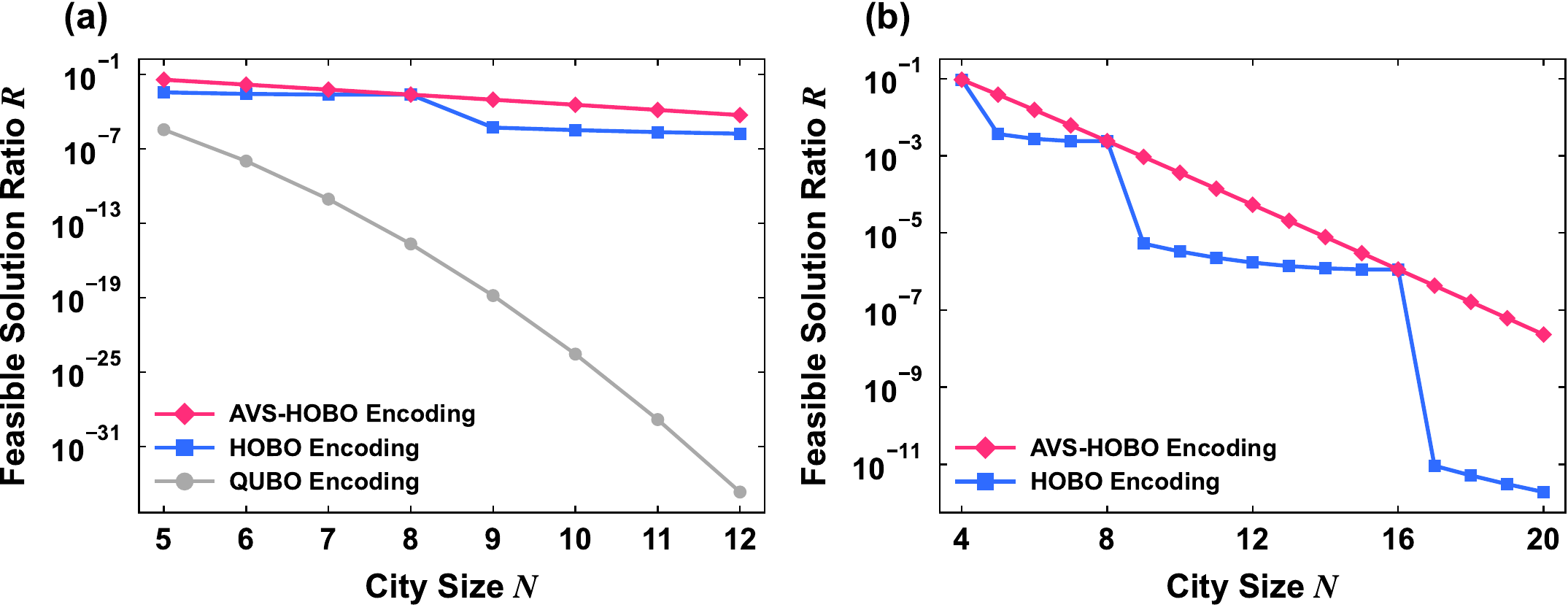}
\caption{Ratio of feasible states in the Hilbert space for different encoding methods, where feasible states are those that satisfy both TSP constraints. (a) Comparison among AVS-HOBO, HOBO, and QUBO encodings for TSP sizes up to 12 cities. The number of feasible states in QUBO differs significantly from that in HOBO, and for 12 cities, the ratio is already close to zero. (b) As the number of cities is further increased, the advantage of AVS-HOBO over HOBO becomes more evident.}
\label{fig3}
\end{figure*}

Therefore, this method eliminates one penalty parameter while further increasing the probability of obtaining feasible solutions. As shown in Fig.~\ref{fig3}(a), compared to QUBO encoding, HOBO encoding yields a considerably larger fraction of feasible states, and this fraction decreases only slowly as the problem size increases. By contrast, as shown in the inset of the figure, when the number of cities reaches 12, the QUBO encoding makes it nearly impossible to locate the optimal solution. Because the present study further scales the problem up to 20 cities, we do not consider QUBO in the subsequent experiments. Compared with the original HOBO encoding, the proposed AVS-HOBO enlarges the feasible solution region while reducing the number of hyperparameters (Fig.~\ref{fig3}(b)) and substantially increases the probability of finding feasible solutions as the problem size increases. We observe that AVS-HOBO and HOBO have the same solution space only when the TSP size is a power of two because every binary string corresponds to a valid city index. A direct comparison of the Hamiltonian formulations reveals that the difference between AVS-HOBO and HOBO is given by

\begin{equation}
\begin{aligned}
\Delta H(b)
&\overset{\triangle}{=}
H^{\mathrm{HOBO}}(b)
-H^{\mathrm{AVS\text{-}HOBO}}(b) \\
&=A_{1}\sum_{t=0}^{N-1}
H_{\mathrm{valid}}(b_{t}).
\end{aligned}
\label{eq11}
\end{equation}

\noindent
accordingly, the number of eliminated terms is
\begin{equation}
\begin{aligned}
\Delta T
&=\left|
\left\{
A_{1}H_{\mathrm{valid}}(b_{t})
\,\middle|\,
t=0,1,\ldots,N-1
\right\}
\right| \\
&=N\in O(N).
\end{aligned}
\label{eq12}
\end{equation}

\begin{figure*}[!t]
\centering
\includegraphics[width=1\textwidth]{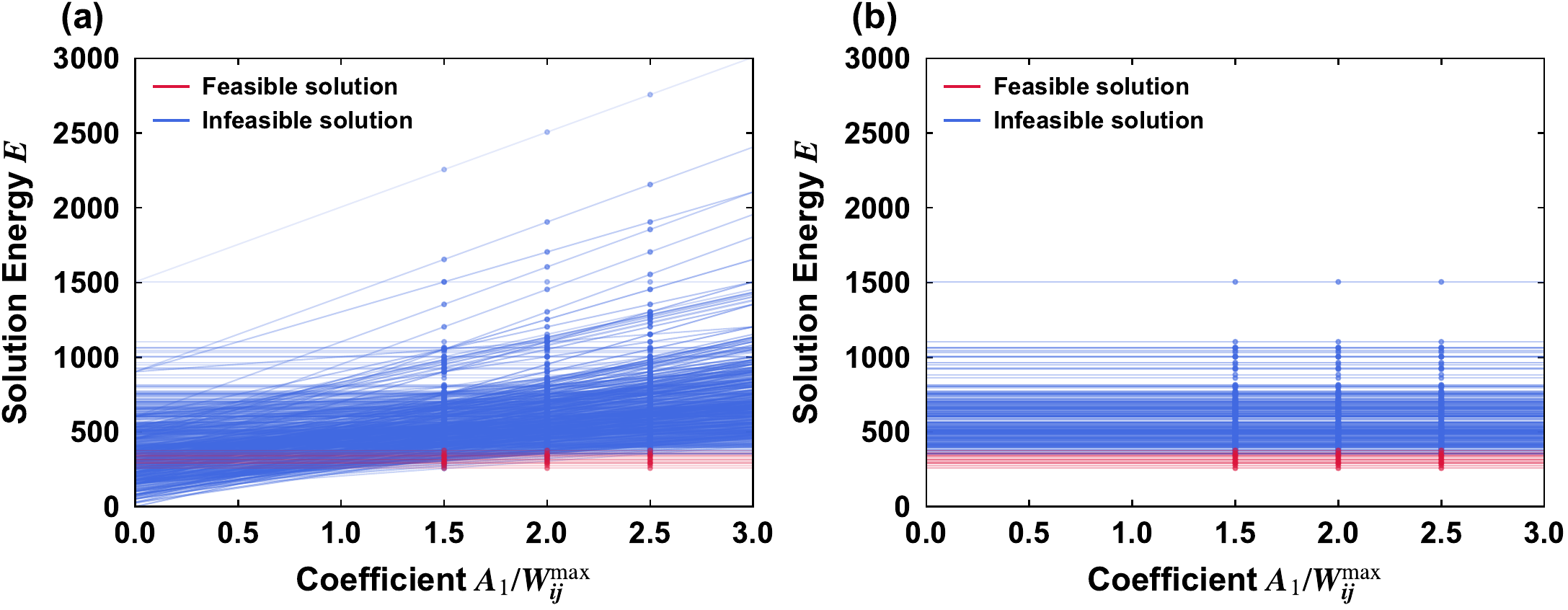}
\caption{Solution energy landscape. (a) Results obtained using the original HOBO method. The horizontal axis shows the coefficient $A_{1}/W_{ij}^{\max}$, which represents the strength of the constraint penalty. Although the original problem formulation contains two constraint terms, here we focus on the optimized term $A_{1}$. As can be seen in the figure, as the constraint penalty becomes stronger, the energies of feasible solutions remain unchanged because they incur no penalty, whereas the energies of infeasible solutions increase accordingly. This enlarges the energy gaps and can make the energy exploration process more difficult and slower. (b) Results obtained using the AVS-HOBO method. Because this method removes the influence of $A_{1}$, the energy of each state remains unchanged regardless of whether the corresponding route is feasible. This can help the energy converge more effectively to some extent.}\label{fig4}
\end{figure*}

\noindent
Therefore, the Hamiltonian is reduced by $\Delta T=N\in O(N)$ terms, consequently reducing the number of penalty hyperparameters by one and eliminating $A_{1}$. However, the introduction of cyclic mapping may also introduce additional overhead. After cyclic mapping, out-of-range binary labels are redefined as valid city indices. As can be observed from Eq.~\eqref{eq6}, multiple binary labels may correspond to the same city index, which can introduce additional penalty contributions in the non-repetition constraint and increase the complexity of the mapped indicator function in the Hamiltonian. Nevertheless, this overhead does not require additional qubits or additional penalty hyperparameters. This study involves numerous different TSP instances. Here, we focus on one representative case: instance 1 of the smallest 5-city problem. For a 5-city TSP, both HOBO and AVS-HOBO require 15 qubits, yielding $2^{15}=32768$ possible quantum states. Even after accounting for duplicate routes, the state space remains enormous. Therefore, we visualize the energies of all states under different values of the constraint coefficient $A_{1}$ as a solution energy landscape \cite{zaborniak2023discrete}, as shown in Fig.~\ref{fig4}(a) and Fig.~\ref{fig4}(b). Fig.~\ref{fig4}(b) also confirms that the cyclic mapping used in AVS-HOBO, which reduces the number of constraints, can improve performance.

\subsection{VQE for optimization}\label{subsec2.5}

Regarding optimizer selection, VQA is among the most promising approaches for NISQ devices. In this study, we adopted one of the best-known methods, VQE. VQE is a hybrid quantum--classical algorithm that searches for the ground state of the target Hamiltonian $\hat{H}$ by minimizing the expectation value
\begin{equation}
E(\theta)=\langle \psi(\theta)\vert \hat{H}\vert \psi(\theta)\rangle.
\label{eq13}
\end{equation}

Here, $\lvert \psi(\theta)\rangle$ is a parameterized quantum state, often referred to as an ansatz. It is implemented using a parameterized quantum circuit whose real-valued parameters $\theta$ determine the prepared quantum state. After the circuit is executed on quantum hardware, the VQE measures the expectation value of the Hamiltonian and feeds the resulting energy back to a classical computer. The classical optimizer updates the parameters to a new set $\theta^{\prime}$, which is sent back to the ansatz to prepare a new state and to be measured again. In this process, the choice of classical optimizer is crucial. In our experiments, we employed NFT \cite{nakanishi2020sequential}, Adam \cite{kingma2014adam}, and CoolMomentum \cite{tsukayama2023coolmomentum} as optimizers and then selected different optimizers depending on the problem settings. For the VQE, this quantum--classical feedback loop is repeated until the energy converges to the ground state.

Although QAOA can be viewed as a problem-inspired ansatz within the broader VQE framework, the term VQE in this study refers to a more general hardware-efficient variational approach. Compared with high-depth QAOA circuits, the hardware-efficient ansatz adopted here can be implemented with relatively shallow circuits, which is advantageous for NISQ devices. Moreover, high-depth QAOA circuits may lead to more complicated energy landscapes with many local minima, where classical optimizers can become trapped \cite{schnaus2024efficient}. In addition, combinatorial optimization problems can be transformed into a Hamiltonian whose energy $E$ encodes the cost function, making the VQE framework a powerful tool for solving such problems.

\section{Results and discussion}\label{sec3}

In this study, we applied the AVS-HOBO method to TSP-related quantum computation for the first time and employed the VQE to compute and compare the resulting performance. In addition to numerical simulations, we collected experimental results on real quantum hardware, specifically the ``ibm\_boston'' backend and the newer ``ibm\_miami'' backend. In the simulations, to mimic the operation of NISQ devices, we employed the Qiskit framework \cite{javadi2024quantum}, and under its noiseless ideal setting, we chose the matrix-product-state (MPS) simulator \cite{bib27}. MPS achieves higher simulation efficiency while consuming less memory. Although a single-layer quantum circuit composed only of $R_{Y}$ gates showed good performance in preliminary tests on small TSP instances, we adopted the EfficientSU2 ansatz following the advice of Qiskit recommendations \cite{bib27,matsuo2023enhancing}. In contrast to circuits that employ only independent $R_{Y}$ gates, this ansatz contains various CNOT gates that generate entanglement. To reduce the number of real-valued parameters $\theta$ and reduce optimization difficulty, we fix part of the circuit to use Y-rotation gates and complement the remaining structure with $R_{X}$ and CNOT gates. For the classical optimizer, we employed NFT \cite{nakanishi2020sequential}, which converges rapidly and is robust against noise.

In the implementation, the number of trainable parameters was set to twice the number of qubits. To ensure convergence for relatively large TSP instances, we initially set the maximum number of iterations to 500 for each instance. However, practically, we observed that when the number of cities increased to 20, 500 iterations were insufficient for reaching a stable solution. After testing larger iteration counts of 600, 700, and 800, we adopted 800 iterations for the 20-city TSP instances. Regarding penalty coefficient $A$, we assumed that different TSP instances require different constraint strengths. Therefore, in the present experiments, we set $A=aW_{ij}^{\max}$, where $W_{ij}^{\max}$ denotes the distance between the two farthest cities in each instance, and $a$ is a constant prefactor, such that the constraint strength is dynamically adapted to each instance. In the following section, we present all residual energies and, based on both simulation and experimental results, compare the original HOBO and AVS-HOBO encodings from three perspectives: the approximation ratio (AR), length ratio (LR), and feasibility ratio (FR).

\subsection{Convergence of the energy}\label{subsec3.1}

\begin{figure*}[!t]
\centering
\includegraphics[width=0.9\textwidth]{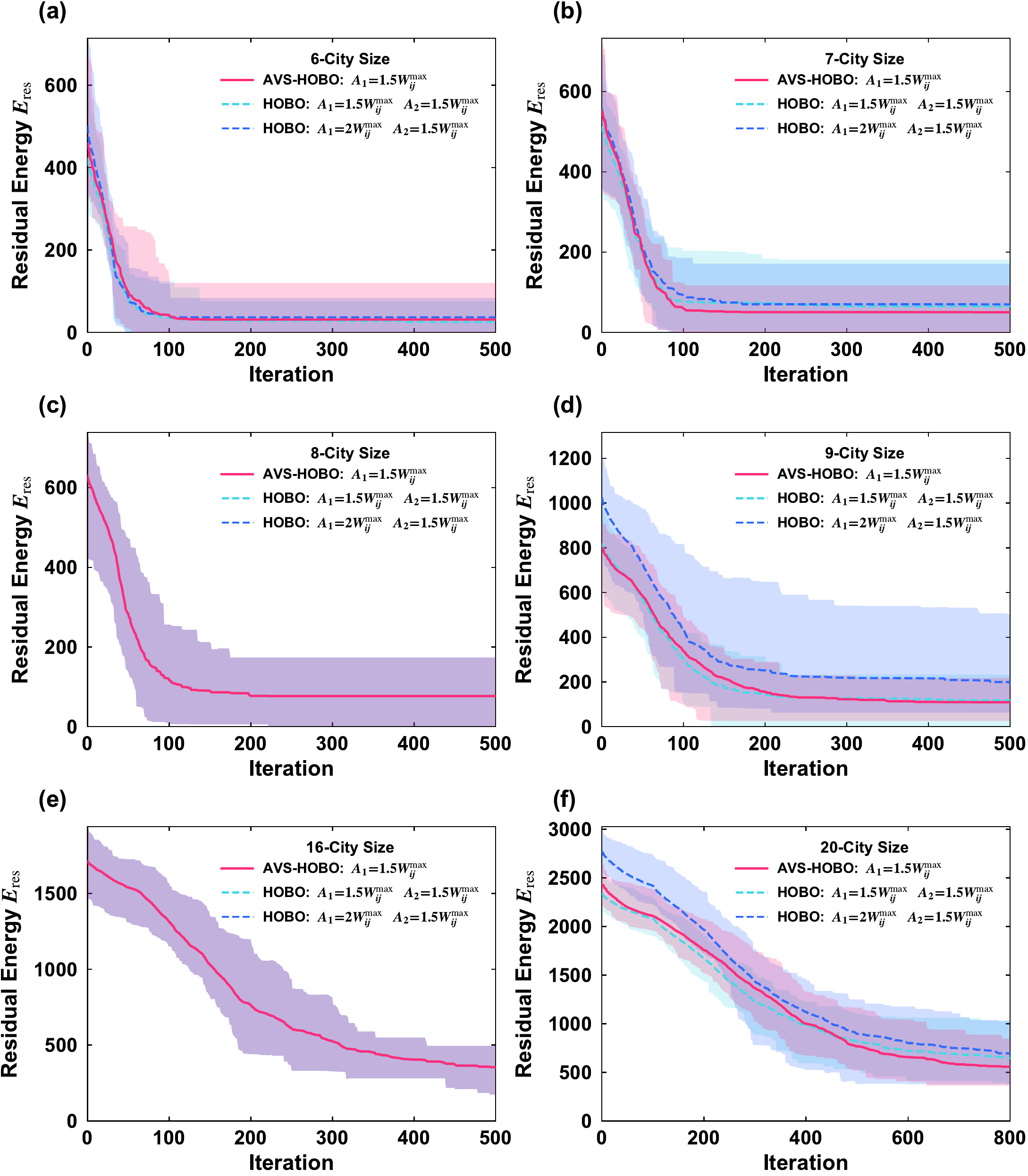}
\caption{Relationship between the residual energy and the iteration number for the TSP. The figure shows the residual energies of 20 different instances under three different conditions for each problem size: (a) 6-city size; (b) 7-city size; (c) 8-city size; (d) 9-city size; (e) 16-city size; and (f) 20-city size. In the figure, both the solid and dashed lines represent the average values over 20 different instances for the corresponding methods, while the shaded area represents the min-max range determined by the maximum and minimum values among the different instances.}\label{fig5}
\end{figure*}

First, we investigate the performance of different methods on the TSP by examining the energy convergence behavior in simulations. Because each TSP instance has its own optimal solution, we introduce the residual energy, defined as
\begin{equation}
E_{\mathrm{res}}=E_{\mathrm{conv}}-E_{\mathrm{opt}},
\label{eq14}
\end{equation}

\noindent
to present the results of all 20 instances in a unified manner. Here, $E_{\mathrm{conv}}$ denotes the energy obtained by the VQE at each iteration, and $E_{\mathrm{opt}}$ is the energy of the optimal solution for the corresponding instance. A residual energy closer to zero indicates that the converged result is closer to the optimum. Fig.~\ref{fig5} shows the relationship between the residual energy and the iteration number for the TSP. For small-scale TSPs, as shown in Fig.~\ref{fig5}(a), the convergence behaviors of the original HOBO and AVS-HOBO are generally comparable. However, as the problem size increases, the advantage of AVS-HOBO gradually becomes apparent. For example, for the 7-city size shown in Fig.~\ref{fig5}(b), the average residual energy after convergence is lower for AVS-HOBO than for HOBO, indicating better convergence performance. As shown in Fig.~\ref{fig5}(c) and Fig.~\ref{fig5}(e), for the 8-city and 16-city sizes, consistent with the previous discussion, the convergence behaviors of the two methods are identical because no invalid states exist. In particular, when $N\geq 9$, as shown in Fig.~\ref{fig5}(d) and Fig.~\ref{fig5}(f), it can be observed that although the initial parameters are randomly set, the HOBO formulation with the stronger constraint $A_{1}=2W_{ij}^{\max}$ tends to start the optimization from a higher-energy region, thereby making convergence more difficult. By contrast, AVS-HOBO, which removes the $A_{1}$-related constraint, can begin the optimization from a lower-energy initial region. The results for the 20-city size indicate that, within the iteration range of 600--800, more AVS-HOBO instances converge to a lower-energy region, suggesting that it can reach a more stable energy state more rapidly. Moreover, it can be clearly seen from the figure that the min-max band of AVS-HOBO is narrower, indicating that the energy evolution is more stable during the convergence process.

\subsection{Approximation ratio of the energy}\label{subsec3.2}

By compiling the TSP into a Hamiltonian and applying the VQE, we searched for the minimum energy of the corresponding Hamiltonian system. To quantify the closeness of the VQE result to the optimal solution, we compared the converged energy with the global optimal energy obtained using a classical solver and defined the AR \cite{qian2023comparative} as
\begin{equation}
\mathrm{AR}=\frac{E_{\mathrm{global}}}{E_{\mathrm{conv}}},
\label{eq15}
\end{equation}

\noindent
where $E_{\mathrm{global}}$ is the optimal energy obtained by a classical algorithm and $E_{\mathrm{conv}}$ is the energy of the converged VQE state. If $\mathrm{AR}=1$, then the quantum result coincides with the classical optimal solution. From the numerical results, we first observe that, for the 8-city TSP, a 3-bit binary string is sufficient to represent exactly eight states; therefore, no invalid city labels appear in the binary encoding. In this case, the original HOBO and AVS-HOBO encodings exhibited identical performance, which is consistent with the theoretical discussion in Section~\ref{sec2}. Using this special case as a reference, we examined both smaller and larger cities to cover a wide range of problem scales. Here, the TSP sizes are chosen as $\mathrm{City}_{\mathrm{node}}\in[5,12]\cup\{16,20\}$. For relatively large TSP instances, a brute-force search on a classical computer becomes infeasible. Therefore, we employed the dedicated TSP algorithm IBM Cplex \cite{bib29} to obtain optimal solutions.

\begin{figure}[t]
\centering
\includegraphics[width=1\columnwidth]{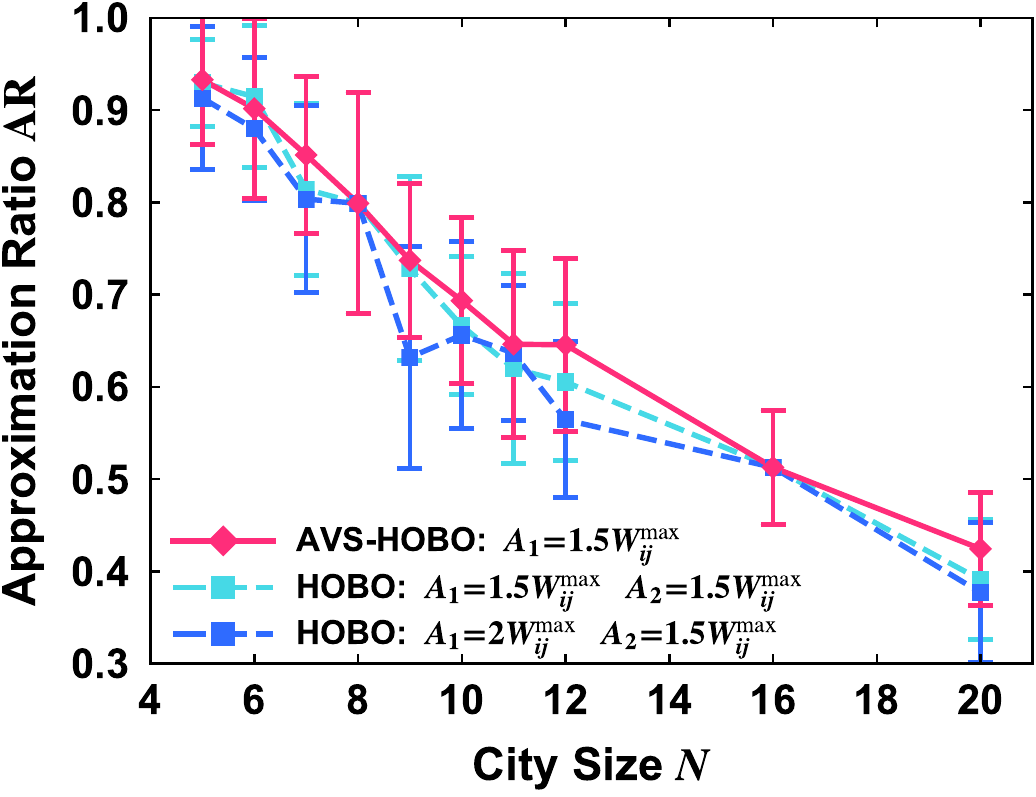}
\caption{Average approximation ratio over 20 instances for each city size in simulations. The results for the 8-- and 16-city sizes are consistent with the theoretical discussion, yielding the same conclusions. Error bars represent standard deviations. The upper and lower bounds of each data set can be seen from the spread, and AVS-HOBO generally outperforms the original method.}\label{fig6}
\end{figure}

To ensure statistical reliability, we generated 20 random instances for each city. For each instance, city coordinates $(x,y)$ are sampled uniformly from the square $[0,100]\times[0,100]$ with different random seeds, and the corresponding TSP is constructed from the Euclidean distances \cite{schnaus2024efficient}. For a fixed city size $N$, we compute the instance-wise AR for all 20 instances, and then take their average. In Fig.~\ref{fig6}, the horizontal axis indicates the city size $N$. For smaller problem sizes, both the original HOBO and AVS-HOBO encodings achieved high AR values close to 1, indicating near-optimal solutions. However, as the city size exceeds $N=8$, the binary encoding switches to a 4-bit representation for each city, and the number of invalid city labels increases. This renders the original HOBO encoding more susceptible to generating invalid city states during VQE optimization. In the 8--12 city range, the AR of the original HOBO encoding decreases noticeably faster than that of AVS-HOBO. At $N=16$, no invalid city indices exist, and the two encodings yield essentially the same performance. For the 20-city TSP, the AVS-HOBO encoding produces converged energies that are closer to the classical optimum, yielding a higher AR than the original HOBO encoding.

\subsection{Length ratio of the TSP}\label{subsec3.3}

Here, we discuss the LR of the TSP \cite{palackal2023quantum}, which serves as another metric for evaluating solution quality. It is defined as
\begin{equation}
\resizebox{0.88\columnwidth}{!}{$
\displaystyle
\mathrm{LR}
=
\frac{\text{optimal TSP path length}}
{\text{average TSP path length for feasible state}}
$}
\label{eq16}
\end{equation}

\begin{figure*}[!t]
\centering
\includegraphics[width=1\textwidth]{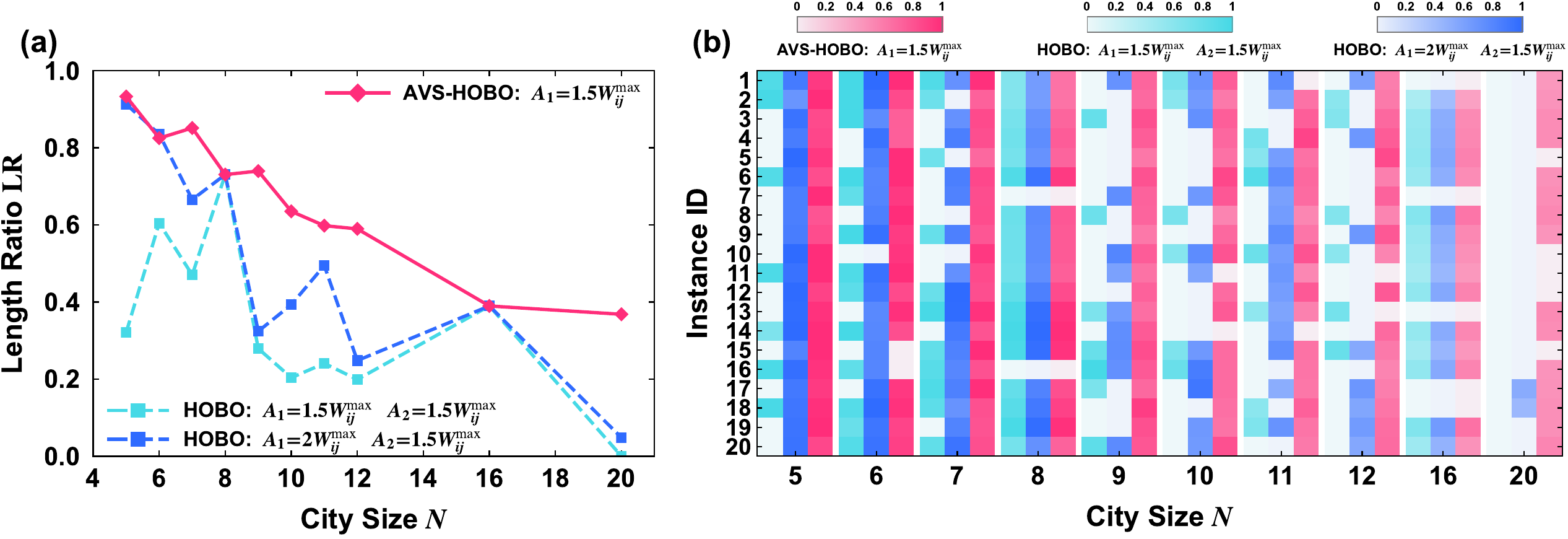}
\caption{Length ratio for the TSP using AVS-HOBO and HOBO. (a) Average LR over 20 instances for each city size in simulations. For the 20-city size, HOBO with $A_{1}=1.5W_{ij}^{\max}$ should, in principle, yield an LR that diverges (tends to infinity) because no feasible route is obtained. However, for ease of visual comparison, we set LR to 0 in this case to indicate the absence of feasible routes. (b) LR of all instances for different city sizes. Instance ID represents the index of each instance, for the same city size, the TSP problem is identical for the same instance. Overall, the comparison shows that, after removing one constraint term, AVS-HOBO achieves better solution quality.}\label{fig7}
\end{figure*}

The optimal TSP path length in the numerator is obtained using a classical TSP solver. In the denominator, the feasible states are those quantum states that satisfy all TSP constraints; thus, they correspond to valid routes. After the VQE optimization essentially converges, the feasible states from the measurement outcomes are extracted, their average path length is computed, and LR is evaluated. As in the previous subsection, LR is averaged over 20 random instances for each city size. As shown in Fig.~\ref{fig7}(a), after one constraint term is removed, AVS-HOBO exhibits clearly higher LR values than the original HOBO encoding, indicating improved solution quality across problems of different difficulty levels. Further comparison of the results for larger problem sizes suggests that, as the problem size increases, the penalty term $A_{1}$ removed in AVS-HOBO has a significant impact on the ability to obtain optimal solutions. This observation indicates that AVS-HOBO can better avoid this sensitivity to the choice of the penalty parameter. The instance-wise distribution of LR is further presented in Fig.~\ref{fig7}(b).

\subsection{Feasibility ratio of the TSP}\label{subsec3.4}

Thus far, we have observed that AVS-HOBO slightly outperforms the original HOBO encoding in terms of convergence and maintains a comparable solution quality. When we consider the FR \cite{schnaus2024efficient}, the advantages become more evident. FR is defined as the fraction of fully feasible routes among all sampled routes after convergence; that is,
\begin{equation}
\mathrm{FR}=\frac{Number_{\mathrm{feasible\ routes}}}{Number_{\mathrm{all\ routes}}}.
\label{eq17}
\end{equation}

\begin{figure*}[!tp]
\centering
\includegraphics[width=1\textwidth]{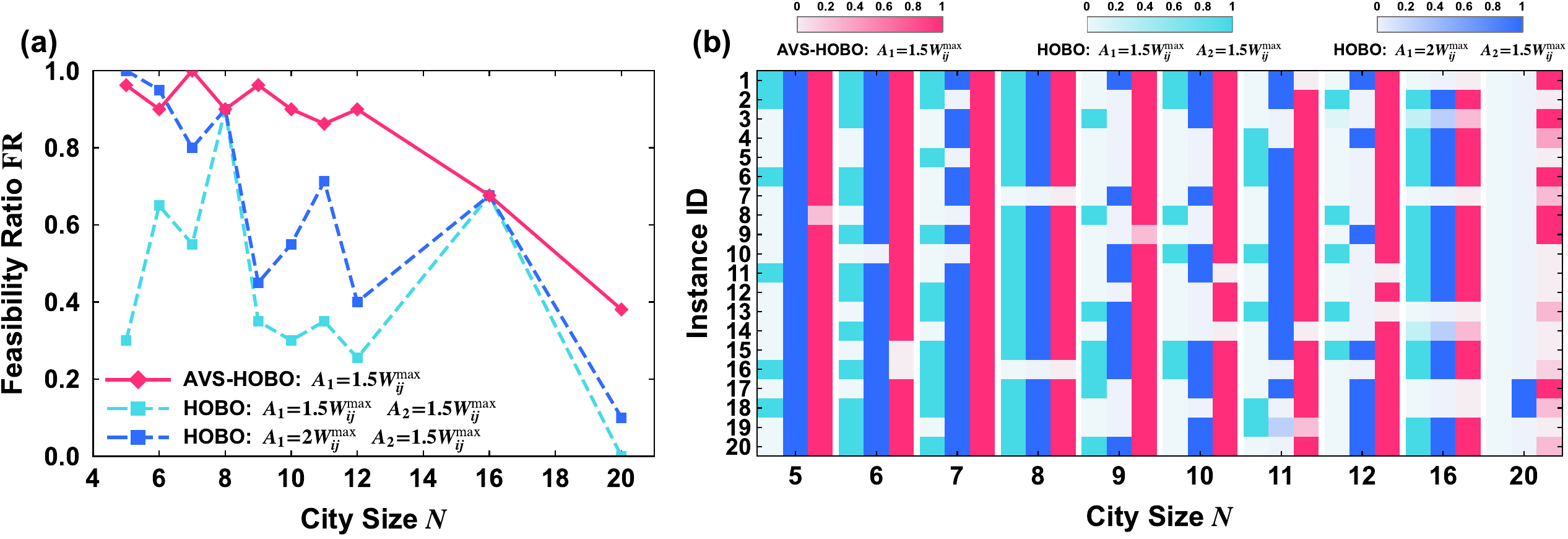}
\caption{Feasibility ratio for the TSP using AVS-HOBO and HOBO. (a) Average FR over 20 instances for each city size in simulations. (b) FR of all instances for different city sizes. Instance ID represents the index of each instance, for the same city size, the TSP problem is identical for the same instance. As the problem becomes more complex, AVS-HOBO exhibits a much higher probability of obtaining feasible solutions than the original method. This is because HOBO must balance the interplay between two constraint terms, and overly large penalty coefficients can instead hinder the energy-exploration process.}\label{fig8}
\end{figure*}

Here, feasible routes are those that satisfy all TSP constraints and thus correspond to valid routes. As shown in Fig.~\ref{fig8}(a), for TSP instances with 5 to 12 cities, AVS-HOBO consistently achieves high FR values, typically between 90\% and 100\%, which are higher than those of HOBO encoding. When the TSP size increases to 16 and 20 cities, the FR of both methods decreases to different extents. In 16 cities, all binary strings correspond to valid city indices; thus, both methods yield the same FR. For the 20-city TSP, the FR of the original HOBO encoding reduces to 0\%--10\%, indicating that feasible routes are rarely obtained, whereas AVS-HOBO maintains an FR of approximately 40\%. HOBO could possibly be improved by tuning the penalty coefficient $A_{2}$, but HOBO encoding requires simultaneous tuning of two penalty parameters, $A_{1}$ and $A_{2}$, which complicates optimization. To further illustrate the variation across instances, Fig.~\ref{fig8}(b) presents the instance-wise distribution of FR.

This aids in interpreting Figs.~\ref{fig6} and \ref{fig8}(a). For example, for the 6-city TSP, HOBO with constraint $A_{1}=1.5W_{ij}^{\max}$ achieves an AR of approximately 0.91, which is slightly higher than that of the AVS-HOBO. However, the corresponding FR was approximately 0.65, indicating that only a small fraction of the 20 instances produced feasible routes. Consequently, the average AR of HOBO is biased by a few feasible samples. When the city size reaches 12, Fig.~\ref{fig6} indicates that AVS-HOBO outperforms HOBO in terms of energy convergence, and Fig.~\ref{fig8}(a) demonstrates that its FR reaches 99.9\%. These observations indicate that AVS-HOBO exhibits stronger stability and a higher probability of obtaining feasible solutions for more complex TSP instances and optimal solutions. By contrast, when numerous invalid states appear in the quantum state distribution of the original HOBO encoding, the FR drops significantly, and the final converged solutions often fail to satisfy the TSP constraints.

\subsection{Experimental results on real quantum hardware}\label{subsec3.5}

Finally, we report the results obtained on real quantum hardware on the IBM Quantum system in Kawasaki. In this set of experiments, we used two chips, namely ``ibm\_boston'' with a Heron r3 processor and ``ibm\_miami'' with a Nighthawk r1 processor. The boston device provides relatively representative, mid-range performance in terms of error mitigation capability and overall hardware quality. By contrast, miami is a newer device. Compared with the more specialized coupling layout of boston, miami connects qubits in a more comprehensive manner, enabling stronger entanglement among qubits \cite{bib27}. The device metrics used in our experiments are summarized as follows. For ibm\_boston, the median two-qubit error is $1.18\times 10^{-3}$, the median readout error is $5.13\times 10^{-3}$, and the circuit layer operations per second (CLOPS) is 340K. This metric represents the number of hardware-native circuit-layer operations that the QPU can execute per second. For ibm\_miami, the median two-qubit error is $2.69\times 10^{-3}$, the median readout error is $1.34\times 10^{-2}$, and the CLOPS is 24K. In the hardware runs, we reused the optimized parameters obtained from the simulations and directly executed the corresponding circuits on real devices \cite{barron2024provable}. Because of limited device access time, the miami experiments were only conducted for city sizes 9, 10, 12, and 20. The resulting trends are consistent with those observed in both the simulations and the boston experiments. This suggests that similar conclusions can be drawn from relatively small problem sizes.

\begin{figure}[!t]
\centering
\includegraphics[width=1\columnwidth]{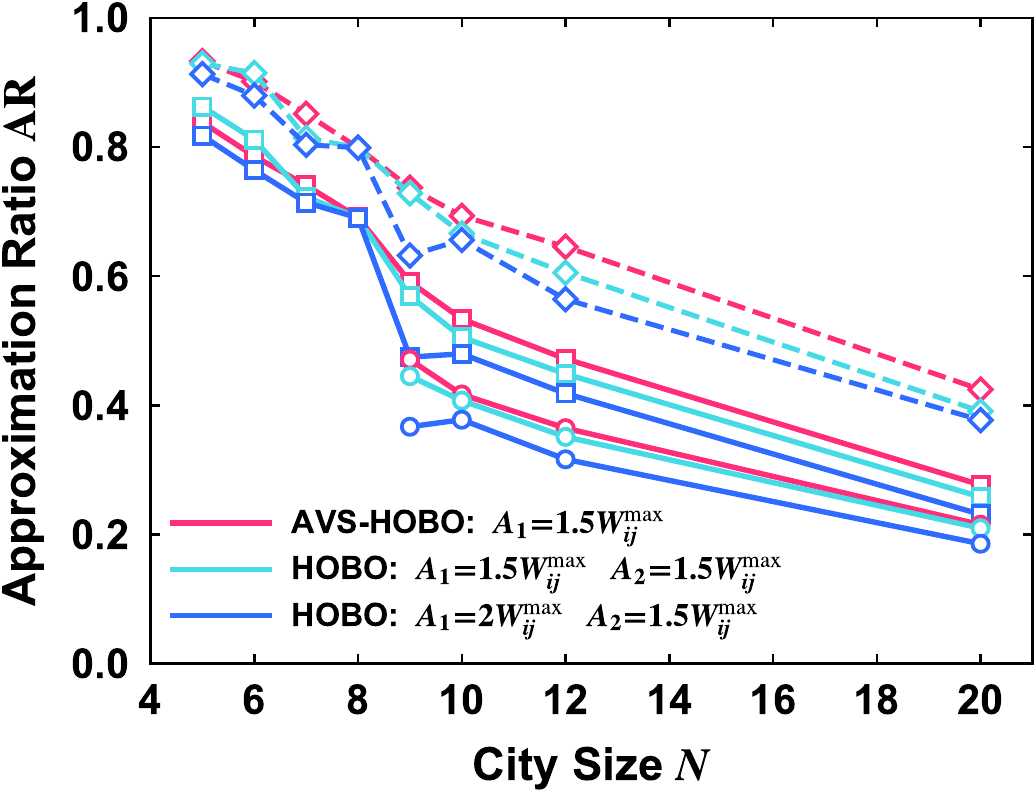}
\caption{Average AR values for different city sizes obtained on real hardware using ibm\_boston (squares) and ibm\_miami (circles), together with the simulation results (dashed lines with diamonds) described above. ibm\_boston uses a Heron r3 processor, whereas ibm\_miami uses Nighthawk r1. The simulation is performed using the MPS simulator. Overall, the AR values obtained on ibm\_boston are lower than the simulation results because of hardware errors, but they are consistently better than those obtained on ibm\_miami.}\label{fig9}
\end{figure}

We continue to analyze the results in terms of AR, LR, and FR. As shown in Fig.~\ref{fig9}, the AR values on boston are reduced by approximately 25\% overall owing to hardware errors. Nevertheless, AR still decreases as the city size increases, and AVS-HOBO consistently outperforms the original HOBO. For miami, AR is lower than that of boston by approximately 22\%. Based on the hardware error messages, we infer that the stronger entanglement makes the circuits more susceptible to errors during execution. This observation also suggests that a fully connected architecture may be more suitable for complex quantum algorithms that contain numerous CNOT gates.

\begin{figure*}[!htbp]
\centering
\includegraphics[width=1\textwidth]{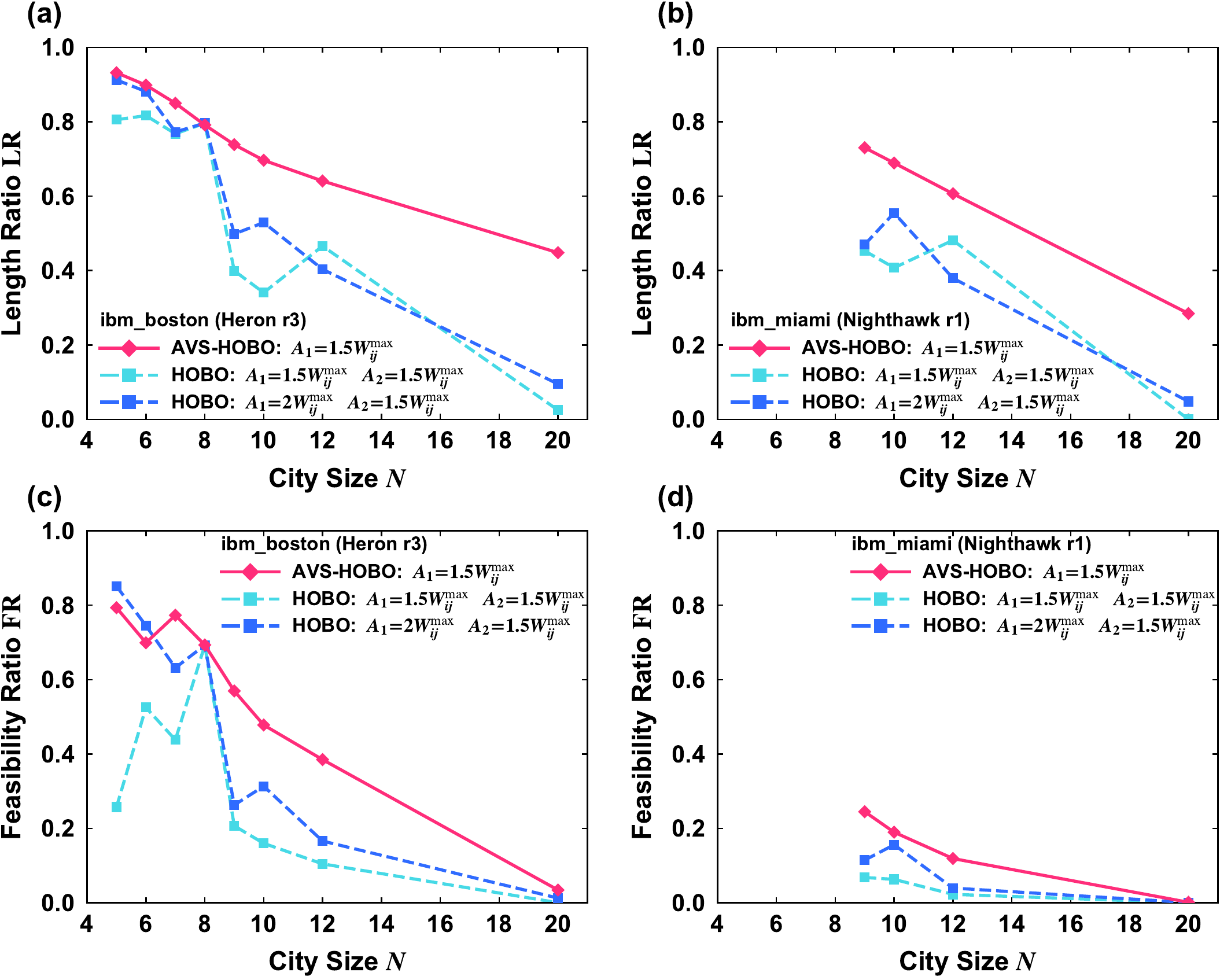}
\caption{Average LR and FR values for different city sizes. (a) and (c) show the results obtained on the real quantum hardware ibm\_boston (processor type: Heron r3), whereas (b) and (d) show those obtained on ibm\_miami (processor type: Nighthawk r1). (a) and (b) present the average LR values, whereas (c) and (d) present the average FR values.}\label{fig10}
\end{figure*}

Fig.~\ref{fig10}(a) and Fig.~\ref{fig10}(b) present the LR values obtained on ibm\_boston and ibm\_miami, respectively. Compared with the simulation results in Fig.~\ref{fig7}, no pronounced degradation is observed. This indicates that the proposed approach can maintain good solution quality in practical computations. Notably, in the 20-city results on boston, HOBO with the constraint $A_{1}=1.5W_{ij}^{\max}$ yields an LR close to 0.5. This behavior arises because LR is computed only from feasible shots, and hardware noise produces very few feasible shots. This effect becomes clear in the subsequent FR analysis. Fig.~\ref{fig10}(c) shows the results obtained on ibm\_boston, whereas Fig.~\ref{fig10}(d) shows those obtained on ibm\_miami. For the 20-city instances, the hardware struggles to find feasible routes. In this case, 100 qubits are already involved, and this poses a highly challenging task for current devices. Overall, the hardware results preserve the same trend as the simulation in Fig.~\ref{fig8}. As the problem becomes more complex, AVS-HOBO still yields substantially more feasible solutions than the original HOBO.

\subsection{Error mitigation on real quantum hardware}\label{subsec3.6}

To further improve the accuracy of experiments on quantum computers, we employed two error-reduction techniques: matrix-free measurement mitigation (Mthree) \cite{chowdhury2024enhancing,lubinski2024quantum} and dynamical decoupling (DD) \cite{ezzell2023dynamical}. Mthree mainly operates during the post-processing stage after measurement and is therefore categorized as an error-mitigation technique. In practical quantum devices, the measured bitstrings may differ from the actual quantum states because of readout noise. Mthree is used to reduce such measurement errors in a scalable manner without requiring a full measurement calibration matrix. By contrast, DD is applied during circuit execution and is more appropriately regarded as an error-suppression technique. DD suppresses errors that accumulate when qubits remain idle by inserting additional control pulses into these idle periods, thereby helping protect the qubits from decoherence and improving the reliability of the measured results.

\begin{figure*}[!htbp]
\centering
\includegraphics[width=1\textwidth]{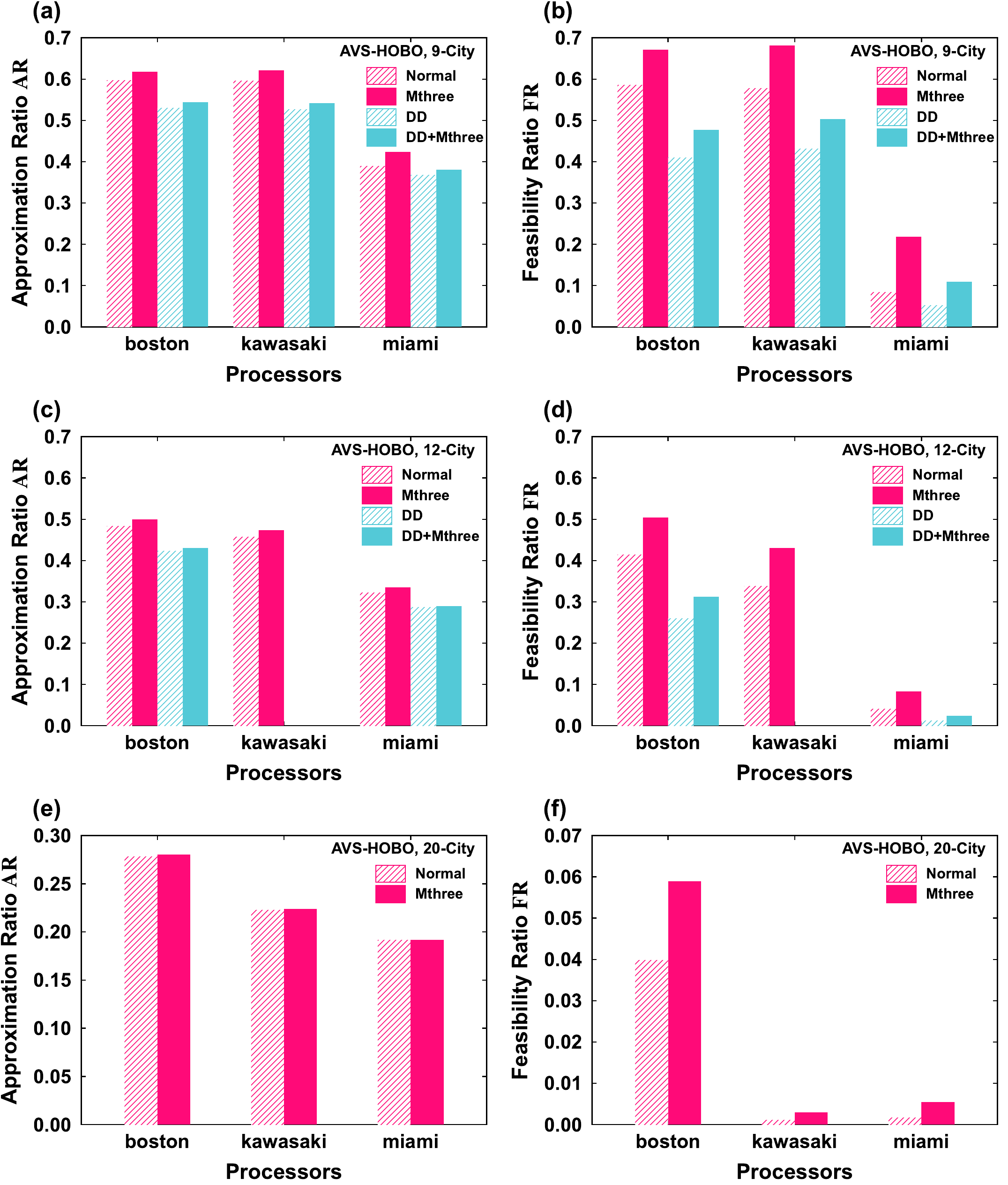}
\caption{Effects of error mitigation strategies on real quantum hardware. (a) and (c) show the approximation ratio (AR), whereas (b) and (d) show the feasibility ratio (FR), for the 9-city and 12-city sizes obtained using the AVS-HOBO method. Specifically, (a) and (b) correspond to the 9-city size, and (c) and (d) correspond to the 12-city size. (e) and (f) show the effect of Mthree on the larger-scale 20-city size, where (e) gives the AR and (f) gives the FR. The horizontal axis represents different quantum processors, and ``Normal'' in the legend denotes the case without any error mitigation. The missing kawasaki data could not be obtained because of limited available runtime. The figure shows that Mthree provides clear performance improvement for the 9-city and 12-city sizes, whereas in the 20-city size its effect on AR is negligible and its improvement in FR is only marginal.}\label{fig11}
\end{figure*}

As shown in Fig.~\ref{fig11}, we analyze the AVS-HOBO method using the 9-city size (Fig.~\ref{fig11}(a) and Fig.~\ref{fig11}(b)) and the 12-city size (Fig.~\ref{fig11}(c) and Fig.~\ref{fig11}(d)) as representative examples, revealing that it exhibits better overall performance. We first examine the convergence behavior in terms of AR. In Fig.~\ref{fig11}(a), boston and kawasaki show similar results. However, as the problem becomes more complex, Fig.~\ref{fig11}(c) shows that boston, based on the Heron r3 processor, exhibits better convergence than kawasaki, which uses the Heron r2 processor, while both perform markedly better than miami, which is based on the Nighthawk r1 processor. Meanwhile, regardless of whether DD is applied, the introduction of Mthree causes a certain degree of improvement in AR. By contrast, after applying DD, all devices show some degradation in convergence performance. Next, we analyze the probability of obtaining feasible solutions, namely FR. For the TSP, feasible solutions are of central importance because a higher probability implies that the quantum computer is more likely to identify the optimal route from the measured results. As shown in Fig.~\ref{fig11}(b) and Fig.~\ref{fig11}(d), Mthree plays a crucial role in real quantum hardware and greatly improves the FR. In contrast, once DD is applied, a clear decrease is observed. This suggests that current quantum computers already possess a certain degree of error-mitigation capability and that additional error suppression during circuit execution is not necessarily beneficial in the present case.

For the missing kawasaki results in the figure, although we attempted to run the experiment on real quantum hardware, the limited available runtime prevented us from obtaining the final results. In the case of miami, the Nighthawk r1 processor adopts a square-lattice architecture. Its CLOPS is only 24K, which is far lower than that of boston (340K) and kawasaki (330K), increasing quantum circuit execution times. This in turn makes the computation more susceptible to error accumulation and leads to degraded performance. Nevertheless, we believe that such an architecture may provide stronger entanglement capability and support more complex workloads in the future.

Finally, we discuss an interesting phenomenon observed for the larger 20-city size. In this case, the number of qubits used reaches 100. As shown in Fig.~\ref{fig11}(e), even after applying Mthree, no improvement in AR is observed. This further confirms that, as the number of qubits increases, the problem becomes more complex and the distribution of measurement outcomes becomes more dispersed. Consequently, the effectiveness of Mthree tends to decrease and may even become insignificant \cite{yang2022efficient}. However, Fig.~\ref{fig11}(f) shows that Mthree still provides a slight improvement in FR, although the gain is very limited. For example, on boston, with 8192 shots, a 2\% improvement corresponds to an increase of approximately 163 feasible shots. This indicates that, for constrained problems, Mthree remains highly useful.

\section{Conclusions}\label{sec4}

This study systematically investigates the performance of VQE for large-scale TSP instances. For the Hamiltonian construction, a HOBO encoding with an efficient binary representation is adopted, and an AVS-HOBO scheme incorporating cyclic mapping is proposed to eliminate one of the constraint terms. Compared to conventional QUBO encoding, HOBO encoding significantly reduces the number of quantum qubits. By analyzing the fraction of TSP-feasible solutions in the Hilbert space, we demonstrate that AVS-HOBO further increases this fraction beyond the already high proportion achieved by the original HOBO encoding. Based on the FR across tested TSP sizes, AVS-HOBO efficiently drives optimization toward states that satisfy all problem constraints. Furthermore, we analyzed the residual energy and compared the AR of the converged energies. We found that AVS-HOBO achieves better convergence than the original HOBO encoding, indicating that it can identify solutions closer to the classical optimum. As the problem size increased, the solution quality of AVS-HOBO is superior to that of HOBO, exhibiting superiority in the 20-city instances. Simultaneously, AVS-HOBO achieved a higher proportion of feasible solutions, suggesting that for large-scale problems with many constraints, where the target Hilbert space becomes highly complex and optimal solutions are difficult to locate, AVS-HOBO provides a more suitable encoding.

However, as the problem becomes larger and more complex, the VQE measurement outcomes contain more spurious components. Performance improvements typically require larger optimization iterations, which significantly increases computational time. Experimental results obtained on real quantum devices also indicate that different processors yield different performance levels and that current quantum chips already possess a certain degree of error-mitigation capability. For practical use on real machines, Mthree demonstrates a strong error-mitigation effect and, for constrained problems, can further increase the probability of finding the optimal solution.

In the future, we plan to apply this encoding scheme to other practically relevant problems. For example, based on our previous study on the feasibility of autonomously optimizing the FCE experimental parameters using gate-based quantum computers \cite{kanezashi2025utility}, we plan to replace the existing model with the present encoding method. Furthermore, we aim to combine AVS-HOBO with CVaR-VQE \cite{chai2023optimal,barkoutsos2020improving} to further improve the convergence speed and the probability of obtaining optimal solutions. Moreover, fully quantum algorithms that have recently attracted increasing attention, such as Grover's search algorithm \cite{sato2025two,liu2025solving,sano2024accelerating,gilliam2021grover}, can be combined with HOBO-type encodings to offload demanding tasks from classical computation and accelerate the overall performance. Additionally, the results obtained on real quantum hardware are particularly intriguing. They demonstrate that these approaches may be feasible on practical NISQ devices, thereby enabling an assessment of the real-world applicability of quantum algorithms to practical problems.

\section*{Abbreviations}
\begin{description}
\item[AR] approximation ratio
\item[AVS-HOBO] all-valid-state higher-order binary optimization
\item[CLOPS] circuit layer operations per second
\item[CVaR-VQE] conditional value-at-risk variational quantum eigensolver
\item[DD] dynamical decoupling
\item[FCE] feedback-controlled electromigration
\item[FR] feasibility ratio
\item[HOBO] higher-order binary optimization
\item[LR] length ratio
\item[MPS] matrix-product-state
\item[Mthree] matrix-free measurement mitigation
\item[NISQ] noisy intermediate-scale quantum
\item[QAOA] quantum approximate optimization algorithm
\item[QUBO] quadratic unconstrained binary optimization
\item[TSP] traveling salesperson problem
\item[VQA] variational quantum algorithms
\item[VQE] variational quantum eigensolver
\end{description}

\section*{Acknowledgements}

The authors express their sincere gratitude to Dr. Atsushi Matsuo and Mr. Toru Imai of IBM Japan for their invaluable discussions and constructive suggestions.

\section*{Declarations}

\begin{itemize}
\item Funding: Not applicable.

\item Data availability: The data that support the findings of this article are not publicly available. The data are available from the authors upon reasonable request.

\item Author contribution: J.W. conceptualized the study, developed the methodology, conducted the experiments, analyzed the data, and drafted the manuscript. T.K. and D.T. provided technical guidance on the experimental design, conducted the experiments, analyzed the data, and contributed to manuscript preparation. K.A. and R.S. provided technical guidance on the experimental design and experimental implementation. J.S., T.S., and H.I. supervised the research, provided critical feedback on the methodology, and contributed to the interpretation of the results. All authors reviewed and approved the final manuscript.
\end{itemize}

\bibliography{reference}

\begin{thebibliography}{40}%
\makeatletter
\providecommand \@ifxundefined [1]{%
 \@ifx{#1\undefined}
}%
\providecommand \@ifnum [1]{%
 \ifnum #1\expandafter \@firstoftwo
 \else \expandafter \@secondoftwo
 \fi
}%
\providecommand \@ifx [1]{%
 \ifx #1\expandafter \@firstoftwo
 \else \expandafter \@secondoftwo
 \fi
}%
\providecommand \natexlab [1]{#1}%
\providecommand \enquote  [1]{``#1''}%
\providecommand \bibnamefont  [1]{#1}%
\providecommand \bibfnamefont [1]{#1}%
\providecommand \citenamefont [1]{#1}%
\providecommand \href@noop [0]{\@secondoftwo}%
\providecommand \href [0]{\begingroup \@sanitize@url \@href}%
\providecommand \@href[1]{\@@startlink{#1}\@@href}%
\providecommand \@@href[1]{\endgroup#1\@@endlink}%
\providecommand \@sanitize@url [0]{\catcode `\\12\catcode `\$12\catcode `\&12\catcode `\#12\catcode `\^12\catcode `\_12\catcode `\%12\relax}%
\providecommand \@@startlink[1]{}%
\providecommand \@@endlink[0]{}%
\providecommand \url  [0]{\begingroup\@sanitize@url \@url }%
\providecommand \@url [1]{\endgroup\@href {#1}{\urlprefix }}%
\providecommand \urlprefix  [0]{URL }%
\providecommand \Eprint [0]{\href }%
\providecommand \doibase [0]{https://doi.org/}%
\providecommand \selectlanguage [0]{\@gobble}%
\providecommand \bibinfo  [0]{\@secondoftwo}%
\providecommand \bibfield  [0]{\@secondoftwo}%
\providecommand \translation [1]{[#1]}%
\providecommand \BibitemOpen [0]{}%
\providecommand \bibitemStop [0]{}%
\providecommand \bibitemNoStop [0]{.\EOS\space}%
\providecommand \EOS [0]{\spacefactor3000\relax}%
\providecommand \BibitemShut  [1]{\csname bibitem#1\endcsname}%
\let\auto@bib@innerbib\@empty
\bibitem [{\citenamefont {Cerezo}\ \emph {et~al.}(2021)\citenamefont {Cerezo}, \citenamefont {Arrasmith}, \citenamefont {Babbush}, \citenamefont {Benjamin}, \citenamefont {Endo}, \citenamefont {Fujii}, \citenamefont {McClean}, \citenamefont {Mitarai}, \citenamefont {Yuan}, \citenamefont {Cincio} \emph {et~al.}}]{cerezo2021variational}%
  \BibitemOpen
  \bibfield  {author} {\bibinfo {author} {\bibfnamefont {M.}~\bibnamefont {Cerezo}}, \bibinfo {author} {\bibfnamefont {A.}~\bibnamefont {Arrasmith}}, \bibinfo {author} {\bibfnamefont {R.}~\bibnamefont {Babbush}}, \bibinfo {author} {\bibfnamefont {S.~C.}\ \bibnamefont {Benjamin}}, \bibinfo {author} {\bibfnamefont {S.}~\bibnamefont {Endo}}, \bibinfo {author} {\bibfnamefont {K.}~\bibnamefont {Fujii}}, \bibinfo {author} {\bibfnamefont {J.~R.}\ \bibnamefont {McClean}}, \bibinfo {author} {\bibfnamefont {K.}~\bibnamefont {Mitarai}}, \bibinfo {author} {\bibfnamefont {X.}~\bibnamefont {Yuan}}, \bibinfo {author} {\bibfnamefont {L.}~\bibnamefont {Cincio}}, \emph {et~al.},\ }\href@noop {} {\bibfield  {journal} {\bibinfo  {journal} {Nature Reviews Physics}\ }\textbf {\bibinfo {volume} {3}},\ \bibinfo {pages} {625} (\bibinfo {year} {2021})}\BibitemShut {NoStop}%
\bibitem [{\citenamefont {Preskill}(2018)}]{preskill2018quantum}%
  \BibitemOpen
  \bibfield  {author} {\bibinfo {author} {\bibfnamefont {J.}~\bibnamefont {Preskill}},\ }\href@noop {} {\bibfield  {journal} {\bibinfo  {journal} {Quantum}\ }\textbf {\bibinfo {volume} {2}},\ \bibinfo {pages} {79} (\bibinfo {year} {2018})}\BibitemShut {NoStop}%
\bibitem [{\citenamefont {Bharti}\ \emph {et~al.}(2022)\citenamefont {Bharti}, \citenamefont {Cervera-Lierta}, \citenamefont {Kyaw}, \citenamefont {Haug}, \citenamefont {Alperin-Lea}, \citenamefont {Anand}, \citenamefont {Degroote}, \citenamefont {Heimonen}, \citenamefont {Kottmann}, \citenamefont {Menke} \emph {et~al.}}]{bharti2022noisy}%
  \BibitemOpen
  \bibfield  {author} {\bibinfo {author} {\bibfnamefont {K.}~\bibnamefont {Bharti}}, \bibinfo {author} {\bibfnamefont {A.}~\bibnamefont {Cervera-Lierta}}, \bibinfo {author} {\bibfnamefont {T.~H.}\ \bibnamefont {Kyaw}}, \bibinfo {author} {\bibfnamefont {T.}~\bibnamefont {Haug}}, \bibinfo {author} {\bibfnamefont {S.}~\bibnamefont {Alperin-Lea}}, \bibinfo {author} {\bibfnamefont {A.}~\bibnamefont {Anand}}, \bibinfo {author} {\bibfnamefont {M.}~\bibnamefont {Degroote}}, \bibinfo {author} {\bibfnamefont {H.}~\bibnamefont {Heimonen}}, \bibinfo {author} {\bibfnamefont {J.~S.}\ \bibnamefont {Kottmann}}, \bibinfo {author} {\bibfnamefont {T.}~\bibnamefont {Menke}}, \emph {et~al.},\ }\href@noop {} {\bibfield  {journal} {\bibinfo  {journal} {Reviews of Modern Physics}\ }\textbf {\bibinfo {volume} {94}},\ \bibinfo {pages} {015004} (\bibinfo {year} {2022})}\BibitemShut {NoStop}%
\bibitem [{\citenamefont {Moll}\ \emph {et~al.}(2018)\citenamefont {Moll}, \citenamefont {Barkoutsos}, \citenamefont {Bishop}, \citenamefont {Chow}, \citenamefont {Cross}, \citenamefont {Egger}, \citenamefont {Filipp}, \citenamefont {Fuhrer}, \citenamefont {Gambetta}, \citenamefont {Ganzhorn} \emph {et~al.}}]{moll2018quantum}%
  \BibitemOpen
  \bibfield  {author} {\bibinfo {author} {\bibfnamefont {N.}~\bibnamefont {Moll}}, \bibinfo {author} {\bibfnamefont {P.}~\bibnamefont {Barkoutsos}}, \bibinfo {author} {\bibfnamefont {L.~S.}\ \bibnamefont {Bishop}}, \bibinfo {author} {\bibfnamefont {J.~M.}\ \bibnamefont {Chow}}, \bibinfo {author} {\bibfnamefont {A.}~\bibnamefont {Cross}}, \bibinfo {author} {\bibfnamefont {D.~J.}\ \bibnamefont {Egger}}, \bibinfo {author} {\bibfnamefont {S.}~\bibnamefont {Filipp}}, \bibinfo {author} {\bibfnamefont {A.}~\bibnamefont {Fuhrer}}, \bibinfo {author} {\bibfnamefont {J.~M.}\ \bibnamefont {Gambetta}}, \bibinfo {author} {\bibfnamefont {M.}~\bibnamefont {Ganzhorn}}, \emph {et~al.},\ }\href@noop {} {\bibfield  {journal} {\bibinfo  {journal} {Quantum Science and Technology}\ }\textbf {\bibinfo {volume} {3}},\ \bibinfo {pages} {030503} (\bibinfo {year} {2018})}\BibitemShut {NoStop}%
\bibitem [{\citenamefont {Perez-Ramirez}(2024)}]{perez2024variational}%
  \BibitemOpen
  \bibfield  {author} {\bibinfo {author} {\bibfnamefont {D.~F.}\ \bibnamefont {Perez-Ramirez}},\ }\href@noop {} {\bibfield  {journal} {\bibinfo  {journal} {arXiv preprint arXiv}\ ,\ \bibinfo {pages} {:2407.06421}} (\bibinfo {year} {2024})}\BibitemShut {NoStop}%
\bibitem [{\citenamefont {Nannicini}(2019)}]{nannicini2019performance}%
  \BibitemOpen
  \bibfield  {author} {\bibinfo {author} {\bibfnamefont {G.}~\bibnamefont {Nannicini}},\ }\href@noop {} {\bibfield  {journal} {\bibinfo  {journal} {Physical Review E}\ }\textbf {\bibinfo {volume} {99}},\ \bibinfo {pages} {013304} (\bibinfo {year} {2019})}\BibitemShut {NoStop}%
\bibitem [{\citenamefont {Glover}\ \emph {et~al.}(2022)\citenamefont {Glover}, \citenamefont {Kochenberger}, \citenamefont {Hennig},\ and\ \citenamefont {Du}}]{glover2022quantum}%
  \BibitemOpen
  \bibfield  {author} {\bibinfo {author} {\bibfnamefont {F.}~\bibnamefont {Glover}}, \bibinfo {author} {\bibfnamefont {G.}~\bibnamefont {Kochenberger}}, \bibinfo {author} {\bibfnamefont {R.}~\bibnamefont {Hennig}},\ and\ \bibinfo {author} {\bibfnamefont {Y.}~\bibnamefont {Du}},\ }\href@noop {} {\bibfield  {journal} {\bibinfo  {journal} {Annals of Operations Research}\ }\textbf {\bibinfo {volume} {314}},\ \bibinfo {pages} {141} (\bibinfo {year} {2022})}\BibitemShut {NoStop}%
\bibitem [{\citenamefont {Lucas}(2014)}]{lucas2014ising}%
  \BibitemOpen
  \bibfield  {author} {\bibinfo {author} {\bibfnamefont {A.}~\bibnamefont {Lucas}},\ }\href@noop {} {\bibfield  {journal} {\bibinfo  {journal} {Frontiers in physics}\ }\textbf {\bibinfo {volume} {2}},\ \bibinfo {pages} {74887} (\bibinfo {year} {2014})}\BibitemShut {NoStop}%
\bibitem [{\citenamefont {Salehi}\ \emph {et~al.}(2022)\citenamefont {Salehi}, \citenamefont {Glos},\ and\ \citenamefont {Miszczak}}]{salehi2022unconstrained}%
  \BibitemOpen
  \bibfield  {author} {\bibinfo {author} {\bibfnamefont {{\"O}.}~\bibnamefont {Salehi}}, \bibinfo {author} {\bibfnamefont {A.}~\bibnamefont {Glos}},\ and\ \bibinfo {author} {\bibfnamefont {J.~A.}\ \bibnamefont {Miszczak}},\ }\href@noop {} {\bibfield  {journal} {\bibinfo  {journal} {Quantum Information Processing}\ }\textbf {\bibinfo {volume} {21}},\ \bibinfo {pages} {67} (\bibinfo {year} {2022})}\BibitemShut {NoStop}%
\bibitem [{\citenamefont {Palackal}\ \emph {et~al.}(2023)\citenamefont {Palackal}, \citenamefont {Poggel}, \citenamefont {Wulff}, \citenamefont {Ehm}, \citenamefont {Lorenz},\ and\ \citenamefont {Mendl}}]{palackal2023quantum}%
  \BibitemOpen
  \bibfield  {author} {\bibinfo {author} {\bibfnamefont {L.}~\bibnamefont {Palackal}}, \bibinfo {author} {\bibfnamefont {B.}~\bibnamefont {Poggel}}, \bibinfo {author} {\bibfnamefont {M.}~\bibnamefont {Wulff}}, \bibinfo {author} {\bibfnamefont {H.}~\bibnamefont {Ehm}}, \bibinfo {author} {\bibfnamefont {J.~M.}\ \bibnamefont {Lorenz}},\ and\ \bibinfo {author} {\bibfnamefont {C.~B.}\ \bibnamefont {Mendl}},\ }\href@noop {} {\bibfield  {journal} {\bibinfo  {journal} {Proceedings of the 2023 IEEE International Conference on Quantum Computing and Engineering (QCE)}\ ,\ \bibinfo {pages} {648}} (\bibinfo {year} {2023})}\BibitemShut {NoStop}%
\bibitem [{\citenamefont {Domino}\ \emph {et~al.}(2022)\citenamefont {Domino}, \citenamefont {Kundu}, \citenamefont {Salehi},\ and\ \citenamefont {Krawiec}}]{domino2022quadratic}%
  \BibitemOpen
  \bibfield  {author} {\bibinfo {author} {\bibfnamefont {K.}~\bibnamefont {Domino}}, \bibinfo {author} {\bibfnamefont {A.}~\bibnamefont {Kundu}}, \bibinfo {author} {\bibfnamefont {{\"O}.}~\bibnamefont {Salehi}},\ and\ \bibinfo {author} {\bibfnamefont {K.}~\bibnamefont {Krawiec}},\ }\href@noop {} {\bibfield  {journal} {\bibinfo  {journal} {Quantum Information Processing}\ }\textbf {\bibinfo {volume} {21}},\ \bibinfo {pages} {337} (\bibinfo {year} {2022})}\BibitemShut {NoStop}%
\bibitem [{\citenamefont {Chai}\ \emph {et~al.}(2023)\citenamefont {Chai}, \citenamefont {Funcke}, \citenamefont {Hartung}, \citenamefont {Jansen}, \citenamefont {K{\"u}hn}, \citenamefont {Stornati},\ and\ \citenamefont {Stollenwerk}}]{chai2023optimal}%
  \BibitemOpen
  \bibfield  {author} {\bibinfo {author} {\bibfnamefont {Y.}~\bibnamefont {Chai}}, \bibinfo {author} {\bibfnamefont {L.}~\bibnamefont {Funcke}}, \bibinfo {author} {\bibfnamefont {T.}~\bibnamefont {Hartung}}, \bibinfo {author} {\bibfnamefont {K.}~\bibnamefont {Jansen}}, \bibinfo {author} {\bibfnamefont {S.}~\bibnamefont {K{\"u}hn}}, \bibinfo {author} {\bibfnamefont {P.}~\bibnamefont {Stornati}},\ and\ \bibinfo {author} {\bibfnamefont {T.}~\bibnamefont {Stollenwerk}},\ }\href@noop {} {\bibfield  {journal} {\bibinfo  {journal} {Physical review applied}\ }\textbf {\bibinfo {volume} {20}},\ \bibinfo {pages} {064025} (\bibinfo {year} {2023})}\BibitemShut {NoStop}%
\bibitem [{\citenamefont {Glos}\ \emph {et~al.}(2022)\citenamefont {Glos}, \citenamefont {Krawiec},\ and\ \citenamefont {Zimbor{\'a}s}}]{glos2022space}%
  \BibitemOpen
  \bibfield  {author} {\bibinfo {author} {\bibfnamefont {A.}~\bibnamefont {Glos}}, \bibinfo {author} {\bibfnamefont {A.}~\bibnamefont {Krawiec}},\ and\ \bibinfo {author} {\bibfnamefont {Z.}~\bibnamefont {Zimbor{\'a}s}},\ }\href@noop {} {\bibfield  {journal} {\bibinfo  {journal} {npj Quantum Information}\ }\textbf {\bibinfo {volume} {8}},\ \bibinfo {pages} {39} (\bibinfo {year} {2022})}\BibitemShut {NoStop}%
\bibitem [{\citenamefont {Schnaus}\ \emph {et~al.}(2024)\citenamefont {Schnaus}, \citenamefont {Palackal}, \citenamefont {Poggel}, \citenamefont {Runge}, \citenamefont {Ehm}, \citenamefont {Lorenz},\ and\ \citenamefont {Mendl}}]{schnaus2024efficient}%
  \BibitemOpen
  \bibfield  {author} {\bibinfo {author} {\bibfnamefont {M.}~\bibnamefont {Schnaus}}, \bibinfo {author} {\bibfnamefont {L.}~\bibnamefont {Palackal}}, \bibinfo {author} {\bibfnamefont {B.}~\bibnamefont {Poggel}}, \bibinfo {author} {\bibfnamefont {X.}~\bibnamefont {Runge}}, \bibinfo {author} {\bibfnamefont {H.}~\bibnamefont {Ehm}}, \bibinfo {author} {\bibfnamefont {J.~M.}\ \bibnamefont {Lorenz}},\ and\ \bibinfo {author} {\bibfnamefont {C.~B.}\ \bibnamefont {Mendl}},\ }\href@noop {} {\bibfield  {journal} {\bibinfo  {journal} {Proceedings of the 2024 IEEE International Conference on Quantum Software (QSW)}\ ,\ \bibinfo {pages} {81}} (\bibinfo {year} {2024})}\BibitemShut {NoStop}%
\bibitem [{\citenamefont {Farhi}\ \emph {et~al.}(2014)\citenamefont {Farhi}, \citenamefont {Goldstone},\ and\ \citenamefont {Gutmann}}]{farhi2014quantum}%
  \BibitemOpen
  \bibfield  {author} {\bibinfo {author} {\bibfnamefont {E.}~\bibnamefont {Farhi}}, \bibinfo {author} {\bibfnamefont {J.}~\bibnamefont {Goldstone}},\ and\ \bibinfo {author} {\bibfnamefont {S.}~\bibnamefont {Gutmann}},\ }\href@noop {} {\bibfield  {journal} {\bibinfo  {journal} {arXiv preprint arXiv}\ ,\ \bibinfo {pages} {:1411.4028}} (\bibinfo {year} {2014})}\BibitemShut {NoStop}%
\bibitem [{\citenamefont {Farhi}\ \emph {et~al.}(2017)\citenamefont {Farhi}, \citenamefont {Goldstone}, \citenamefont {Gutmann},\ and\ \citenamefont {Neven}}]{farhi2017quantum}%
  \BibitemOpen
  \bibfield  {author} {\bibinfo {author} {\bibfnamefont {E.}~\bibnamefont {Farhi}}, \bibinfo {author} {\bibfnamefont {J.}~\bibnamefont {Goldstone}}, \bibinfo {author} {\bibfnamefont {S.}~\bibnamefont {Gutmann}},\ and\ \bibinfo {author} {\bibfnamefont {H.}~\bibnamefont {Neven}},\ }\href@noop {} {\bibfield  {journal} {\bibinfo  {journal} {arXiv preprint arXiv}\ ,\ \bibinfo {pages} {:1703.06199}} (\bibinfo {year} {2017})}\BibitemShut {NoStop}%
\bibitem [{\citenamefont {Qian}\ \emph {et~al.}(2023)\citenamefont {Qian}, \citenamefont {Basili}, \citenamefont {Eshaghian-Wilner}, \citenamefont {Khokhar}, \citenamefont {Luecke},\ and\ \citenamefont {Vary}}]{qian2023comparative}%
  \BibitemOpen
  \bibfield  {author} {\bibinfo {author} {\bibfnamefont {W.}~\bibnamefont {Qian}}, \bibinfo {author} {\bibfnamefont {R.~A.}\ \bibnamefont {Basili}}, \bibinfo {author} {\bibfnamefont {M.~M.}\ \bibnamefont {Eshaghian-Wilner}}, \bibinfo {author} {\bibfnamefont {A.}~\bibnamefont {Khokhar}}, \bibinfo {author} {\bibfnamefont {G.}~\bibnamefont {Luecke}},\ and\ \bibinfo {author} {\bibfnamefont {J.~P.}\ \bibnamefont {Vary}},\ }\href@noop {} {\bibfield  {journal} {\bibinfo  {journal} {Entropy}\ }\textbf {\bibinfo {volume} {25}},\ \bibinfo {pages} {1238} (\bibinfo {year} {2023})}\BibitemShut {NoStop}%
\bibitem [{\citenamefont {Peruzzo}\ \emph {et~al.}(2014)\citenamefont {Peruzzo}, \citenamefont {McClean}, \citenamefont {Shadbolt}, \citenamefont {Yung}, \citenamefont {Zhou}, \citenamefont {Love}, \citenamefont {Aspuru-Guzik},\ and\ \citenamefont {O’brien}}]{peruzzo2014variational}%
  \BibitemOpen
  \bibfield  {author} {\bibinfo {author} {\bibfnamefont {A.}~\bibnamefont {Peruzzo}}, \bibinfo {author} {\bibfnamefont {J.}~\bibnamefont {McClean}}, \bibinfo {author} {\bibfnamefont {P.}~\bibnamefont {Shadbolt}}, \bibinfo {author} {\bibfnamefont {M.-H.}\ \bibnamefont {Yung}}, \bibinfo {author} {\bibfnamefont {X.-Q.}\ \bibnamefont {Zhou}}, \bibinfo {author} {\bibfnamefont {P.~J.}\ \bibnamefont {Love}}, \bibinfo {author} {\bibfnamefont {A.}~\bibnamefont {Aspuru-Guzik}},\ and\ \bibinfo {author} {\bibfnamefont {J.~L.}\ \bibnamefont {O’brien}},\ }\href@noop {} {\bibfield  {journal} {\bibinfo  {journal} {Nature communications}\ }\textbf {\bibinfo {volume} {5}},\ \bibinfo {pages} {4213} (\bibinfo {year} {2014})}\BibitemShut {NoStop}%
\bibitem [{\citenamefont {Tilly}\ \emph {et~al.}(2022)\citenamefont {Tilly}, \citenamefont {Chen}, \citenamefont {Cao}, \citenamefont {Picozzi}, \citenamefont {Setia}, \citenamefont {Li}, \citenamefont {Grant}, \citenamefont {Wossnig}, \citenamefont {Rungger}, \citenamefont {Booth} \emph {et~al.}}]{tilly2022variational}%
  \BibitemOpen
  \bibfield  {author} {\bibinfo {author} {\bibfnamefont {J.}~\bibnamefont {Tilly}}, \bibinfo {author} {\bibfnamefont {H.}~\bibnamefont {Chen}}, \bibinfo {author} {\bibfnamefont {S.}~\bibnamefont {Cao}}, \bibinfo {author} {\bibfnamefont {D.}~\bibnamefont {Picozzi}}, \bibinfo {author} {\bibfnamefont {K.}~\bibnamefont {Setia}}, \bibinfo {author} {\bibfnamefont {Y.}~\bibnamefont {Li}}, \bibinfo {author} {\bibfnamefont {E.}~\bibnamefont {Grant}}, \bibinfo {author} {\bibfnamefont {L.}~\bibnamefont {Wossnig}}, \bibinfo {author} {\bibfnamefont {I.}~\bibnamefont {Rungger}}, \bibinfo {author} {\bibfnamefont {G.~H.}\ \bibnamefont {Booth}}, \emph {et~al.},\ }\href@noop {} {\bibfield  {journal} {\bibinfo  {journal} {Physics Reports}\ }\textbf {\bibinfo {volume} {986}},\ \bibinfo {pages} {1} (\bibinfo {year} {2022})}\BibitemShut {NoStop}%
\bibitem [{\citenamefont {McClean}\ \emph {et~al.}(2016)\citenamefont {McClean}, \citenamefont {Romero}, \citenamefont {Babbush},\ and\ \citenamefont {Aspuru-Guzik}}]{mcclean2016theory}%
  \BibitemOpen
  \bibfield  {author} {\bibinfo {author} {\bibfnamefont {J.~R.}\ \bibnamefont {McClean}}, \bibinfo {author} {\bibfnamefont {J.}~\bibnamefont {Romero}}, \bibinfo {author} {\bibfnamefont {R.}~\bibnamefont {Babbush}},\ and\ \bibinfo {author} {\bibfnamefont {A.}~\bibnamefont {Aspuru-Guzik}},\ }\href@noop {} {\bibfield  {journal} {\bibinfo  {journal} {New Journal of Physics}\ }\textbf {\bibinfo {volume} {18}},\ \bibinfo {pages} {023023} (\bibinfo {year} {2016})}\BibitemShut {NoStop}%
\bibitem [{\citenamefont {Tsukayama}\ \emph {et~al.}(2025)\citenamefont {Tsukayama}, \citenamefont {Shirakashi}, \citenamefont {Shibuya},\ and\ \citenamefont {Imai}}]{tsukayama2025enhancing}%
  \BibitemOpen
  \bibfield  {author} {\bibinfo {author} {\bibfnamefont {D.}~\bibnamefont {Tsukayama}}, \bibinfo {author} {\bibfnamefont {J.-i.}\ \bibnamefont {Shirakashi}}, \bibinfo {author} {\bibfnamefont {T.}~\bibnamefont {Shibuya}},\ and\ \bibinfo {author} {\bibfnamefont {H.}~\bibnamefont {Imai}},\ }\href@noop {} {\bibfield  {journal} {\bibinfo  {journal} {AIP Advances}\ }\textbf {\bibinfo {volume} {15}} (\bibinfo {year} {2025})}\BibitemShut {NoStop}%
\bibitem [{\citenamefont {Zaborniak}\ and\ \citenamefont {Stege}(2023)}]{zaborniak2023discrete}%
  \BibitemOpen
  \bibfield  {author} {\bibinfo {author} {\bibfnamefont {T.}~\bibnamefont {Zaborniak}}\ and\ \bibinfo {author} {\bibfnamefont {U.}~\bibnamefont {Stege}},\ }\href@noop {} {\bibfield  {journal} {\bibinfo  {journal} {arXiv preprint arXiv}\ ,\ \bibinfo {pages} {:2305.00568}} (\bibinfo {year} {2023})}\BibitemShut {NoStop}%
\bibitem [{\citenamefont {Nakanishi}\ \emph {et~al.}(2020)\citenamefont {Nakanishi}, \citenamefont {Fujii},\ and\ \citenamefont {Todo}}]{nakanishi2020sequential}%
  \BibitemOpen
  \bibfield  {author} {\bibinfo {author} {\bibfnamefont {K.~M.}\ \bibnamefont {Nakanishi}}, \bibinfo {author} {\bibfnamefont {K.}~\bibnamefont {Fujii}},\ and\ \bibinfo {author} {\bibfnamefont {S.}~\bibnamefont {Todo}},\ }\href@noop {} {\bibfield  {journal} {\bibinfo  {journal} {Physical Review Research}\ }\textbf {\bibinfo {volume} {2}},\ \bibinfo {pages} {043158} (\bibinfo {year} {2020})}\BibitemShut {NoStop}%
\bibitem [{\citenamefont {Kingma}\ and\ \citenamefont {Ba}(2014)}]{kingma2014adam}%
  \BibitemOpen
  \bibfield  {author} {\bibinfo {author} {\bibfnamefont {D.~P.}\ \bibnamefont {Kingma}}\ and\ \bibinfo {author} {\bibfnamefont {J.}~\bibnamefont {Ba}},\ }\href@noop {} {\bibfield  {journal} {\bibinfo  {journal} {arXiv preprint arXiv}\ ,\ \bibinfo {pages} {:1412.6980}} (\bibinfo {year} {2014})}\BibitemShut {NoStop}%
\bibitem [{\citenamefont {Tsukayama}\ \emph {et~al.}(2023)\citenamefont {Tsukayama}, \citenamefont {Shirakashi},\ and\ \citenamefont {Imai}}]{tsukayama2023coolmomentum}%
  \BibitemOpen
  \bibfield  {author} {\bibinfo {author} {\bibfnamefont {D.}~\bibnamefont {Tsukayama}}, \bibinfo {author} {\bibfnamefont {J.-i.}\ \bibnamefont {Shirakashi}},\ and\ \bibinfo {author} {\bibfnamefont {H.}~\bibnamefont {Imai}},\ }\href@noop {} {\bibfield  {journal} {\bibinfo  {journal} {Japanese Journal of Applied Physics}\ }\textbf {\bibinfo {volume} {62}},\ \bibinfo {pages} {088003} (\bibinfo {year} {2023})}\BibitemShut {NoStop}%
\bibitem [{\citenamefont {Javadi-Abhari}\ \emph {et~al.}(2024)\citenamefont {Javadi-Abhari}, \citenamefont {Treinish}, \citenamefont {Krsulich}, \citenamefont {Wood}, \citenamefont {Lishman}, \citenamefont {Gacon}, \citenamefont {Martiel}, \citenamefont {Nation}, \citenamefont {Bishop}, \citenamefont {Cross} \emph {et~al.}}]{javadi2024quantum}%
  \BibitemOpen
  \bibfield  {author} {\bibinfo {author} {\bibfnamefont {A.}~\bibnamefont {Javadi-Abhari}}, \bibinfo {author} {\bibfnamefont {M.}~\bibnamefont {Treinish}}, \bibinfo {author} {\bibfnamefont {K.}~\bibnamefont {Krsulich}}, \bibinfo {author} {\bibfnamefont {C.~J.}\ \bibnamefont {Wood}}, \bibinfo {author} {\bibfnamefont {J.}~\bibnamefont {Lishman}}, \bibinfo {author} {\bibfnamefont {J.}~\bibnamefont {Gacon}}, \bibinfo {author} {\bibfnamefont {S.}~\bibnamefont {Martiel}}, \bibinfo {author} {\bibfnamefont {P.~D.}\ \bibnamefont {Nation}}, \bibinfo {author} {\bibfnamefont {L.~S.}\ \bibnamefont {Bishop}}, \bibinfo {author} {\bibfnamefont {A.~W.}\ \bibnamefont {Cross}}, \emph {et~al.},\ }\href@noop {} {\bibfield  {journal} {\bibinfo  {journal} {arXiv preprint arXiv}\ ,\ \bibinfo {pages} {:2405.08810}} (\bibinfo {year} {2024})}\BibitemShut {NoStop}%
\bibitem [{\citenamefont {{IBM}}(2026)}]{bib27}%
  \BibitemOpen
  \bibfield  {author} {\bibinfo {author} {\bibnamefont {{IBM}}},\ }\href@noop {} {\bibinfo {title} {Ibm quantum platform}} (\bibinfo {year} {2026}),\ \bibinfo {note} {\url{https://quantum-computing.ibm.com/}.}\BibitemShut {Stop}%
\bibitem [{\citenamefont {Matsuo}\ \emph {et~al.}(2023)\citenamefont {Matsuo}, \citenamefont {Suzuki}, \citenamefont {Hamamura},\ and\ \citenamefont {Yamashita}}]{matsuo2023enhancing}%
  \BibitemOpen
  \bibfield  {author} {\bibinfo {author} {\bibfnamefont {A.}~\bibnamefont {Matsuo}}, \bibinfo {author} {\bibfnamefont {Y.}~\bibnamefont {Suzuki}}, \bibinfo {author} {\bibfnamefont {I.}~\bibnamefont {Hamamura}},\ and\ \bibinfo {author} {\bibfnamefont {S.}~\bibnamefont {Yamashita}},\ }\href@noop {} {\bibfield  {journal} {\bibinfo  {journal} {IEICE TRANSACTIONS on Information and Systems}\ }\textbf {\bibinfo {volume} {106}},\ \bibinfo {pages} {1772} (\bibinfo {year} {2023})}\BibitemShut {NoStop}%
\bibitem [{\citenamefont {{IBM Corporation}}(2026)}]{bib29}%
  \BibitemOpen
  \bibfield  {author} {\bibinfo {author} {\bibnamefont {{IBM Corporation}}},\ }\href@noop {} {\bibinfo {title} {Ibm ilog cplex optimization studio user's manual for cplex}} (\bibinfo {year} {2026}),\ \bibinfo {note} {\url{https://www.ibm.com/docs/en/icos/22.1.1?topic=optimizers-users-manual-cplex}.}\BibitemShut {Stop}%
\bibitem [{\citenamefont {Barron}\ \emph {et~al.}(2024)\citenamefont {Barron}, \citenamefont {Egger}, \citenamefont {Pelofske}, \citenamefont {B{\"a}rtschi}, \citenamefont {Eidenbenz}, \citenamefont {Lehmkuehler},\ and\ \citenamefont {Woerner}}]{barron2024provable}%
  \BibitemOpen
  \bibfield  {author} {\bibinfo {author} {\bibfnamefont {S.~V.}\ \bibnamefont {Barron}}, \bibinfo {author} {\bibfnamefont {D.~J.}\ \bibnamefont {Egger}}, \bibinfo {author} {\bibfnamefont {E.}~\bibnamefont {Pelofske}}, \bibinfo {author} {\bibfnamefont {A.}~\bibnamefont {B{\"a}rtschi}}, \bibinfo {author} {\bibfnamefont {S.}~\bibnamefont {Eidenbenz}}, \bibinfo {author} {\bibfnamefont {M.}~\bibnamefont {Lehmkuehler}},\ and\ \bibinfo {author} {\bibfnamefont {S.}~\bibnamefont {Woerner}},\ }\href@noop {} {\bibfield  {journal} {\bibinfo  {journal} {Nature Computational Science}\ }\textbf {\bibinfo {volume} {4}},\ \bibinfo {pages} {865} (\bibinfo {year} {2024})}\BibitemShut {NoStop}%
\bibitem [{\citenamefont {Chowdhury}\ \emph {et~al.}(2024)\citenamefont {Chowdhury}, \citenamefont {Yu}, \citenamefont {Shamim}, \citenamefont {Kabir},\ and\ \citenamefont {Sufian}}]{chowdhury2024enhancing}%
  \BibitemOpen
  \bibfield  {author} {\bibinfo {author} {\bibfnamefont {T.~A.}\ \bibnamefont {Chowdhury}}, \bibinfo {author} {\bibfnamefont {K.}~\bibnamefont {Yu}}, \bibinfo {author} {\bibfnamefont {M.~A.}\ \bibnamefont {Shamim}}, \bibinfo {author} {\bibfnamefont {M.}~\bibnamefont {Kabir}},\ and\ \bibinfo {author} {\bibfnamefont {R.~S.}\ \bibnamefont {Sufian}},\ }\href@noop {} {\bibfield  {journal} {\bibinfo  {journal} {Physical Review Research}\ }\textbf {\bibinfo {volume} {6}},\ \bibinfo {pages} {033107} (\bibinfo {year} {2024})}\BibitemShut {NoStop}%
\bibitem [{\citenamefont {Lubinski}\ \emph {et~al.}(2024)\citenamefont {Lubinski}, \citenamefont {Goings}, \citenamefont {Mayer}, \citenamefont {Johri}, \citenamefont {Reddy}, \citenamefont {Mehta}, \citenamefont {Bhatia}, \citenamefont {Rappaport}, \citenamefont {Mills}, \citenamefont {Baldwin} \emph {et~al.}}]{lubinski2024quantum}%
  \BibitemOpen
  \bibfield  {author} {\bibinfo {author} {\bibfnamefont {T.}~\bibnamefont {Lubinski}}, \bibinfo {author} {\bibfnamefont {J.~J.}\ \bibnamefont {Goings}}, \bibinfo {author} {\bibfnamefont {K.}~\bibnamefont {Mayer}}, \bibinfo {author} {\bibfnamefont {S.}~\bibnamefont {Johri}}, \bibinfo {author} {\bibfnamefont {N.}~\bibnamefont {Reddy}}, \bibinfo {author} {\bibfnamefont {A.}~\bibnamefont {Mehta}}, \bibinfo {author} {\bibfnamefont {N.}~\bibnamefont {Bhatia}}, \bibinfo {author} {\bibfnamefont {S.}~\bibnamefont {Rappaport}}, \bibinfo {author} {\bibfnamefont {D.}~\bibnamefont {Mills}}, \bibinfo {author} {\bibfnamefont {C.~H.}\ \bibnamefont {Baldwin}}, \emph {et~al.},\ }\href@noop {} {\bibfield  {journal} {\bibinfo  {journal} {arXiv preprint arXiv}\ ,\ \bibinfo {pages} {:2402.08985}} (\bibinfo {year} {2024})}\BibitemShut {NoStop}%
\bibitem [{\citenamefont {Ezzell}\ \emph {et~al.}(2023)\citenamefont {Ezzell}, \citenamefont {Pokharel}, \citenamefont {Tewala}, \citenamefont {Quiroz},\ and\ \citenamefont {Lidar}}]{ezzell2023dynamical}%
  \BibitemOpen
  \bibfield  {author} {\bibinfo {author} {\bibfnamefont {N.}~\bibnamefont {Ezzell}}, \bibinfo {author} {\bibfnamefont {B.}~\bibnamefont {Pokharel}}, \bibinfo {author} {\bibfnamefont {L.}~\bibnamefont {Tewala}}, \bibinfo {author} {\bibfnamefont {G.}~\bibnamefont {Quiroz}},\ and\ \bibinfo {author} {\bibfnamefont {D.~A.}\ \bibnamefont {Lidar}},\ }\href@noop {} {\bibfield  {journal} {\bibinfo  {journal} {Physical Review Applied}\ }\textbf {\bibinfo {volume} {20}},\ \bibinfo {pages} {064027} (\bibinfo {year} {2023})}\BibitemShut {NoStop}%
\bibitem [{\citenamefont {Yang}\ \emph {et~al.}(2022)\citenamefont {Yang}, \citenamefont {Raymond},\ and\ \citenamefont {Uno}}]{yang2022efficient}%
  \BibitemOpen
  \bibfield  {author} {\bibinfo {author} {\bibfnamefont {B.}~\bibnamefont {Yang}}, \bibinfo {author} {\bibfnamefont {R.}~\bibnamefont {Raymond}},\ and\ \bibinfo {author} {\bibfnamefont {S.}~\bibnamefont {Uno}},\ }\href@noop {} {\bibfield  {journal} {\bibinfo  {journal} {Physical Review A}\ }\textbf {\bibinfo {volume} {106}},\ \bibinfo {pages} {012423} (\bibinfo {year} {2022})}\BibitemShut {NoStop}%
\bibitem [{\citenamefont {Kanezashi}\ \emph {et~al.}(2025)\citenamefont {Kanezashi}, \citenamefont {Tsukayama}, \citenamefont {Shirakashi}, \citenamefont {Shibuya},\ and\ \citenamefont {Imai}}]{kanezashi2025utility}%
  \BibitemOpen
  \bibfield  {author} {\bibinfo {author} {\bibfnamefont {T.}~\bibnamefont {Kanezashi}}, \bibinfo {author} {\bibfnamefont {D.}~\bibnamefont {Tsukayama}}, \bibinfo {author} {\bibfnamefont {J.-i.}\ \bibnamefont {Shirakashi}}, \bibinfo {author} {\bibfnamefont {T.}~\bibnamefont {Shibuya}},\ and\ \bibinfo {author} {\bibfnamefont {H.}~\bibnamefont {Imai}},\ }\href@noop {} {\bibfield  {journal} {\bibinfo  {journal} {Applied Physics Express}\ }\textbf {\bibinfo {volume} {18}},\ \bibinfo {pages} {047001} (\bibinfo {year} {2025})}\BibitemShut {NoStop}%
\bibitem [{\citenamefont {Barkoutsos}\ \emph {et~al.}(2020)\citenamefont {Barkoutsos}, \citenamefont {Nannicini}, \citenamefont {Robert}, \citenamefont {Tavernelli},\ and\ \citenamefont {Woerner}}]{barkoutsos2020improving}%
  \BibitemOpen
  \bibfield  {author} {\bibinfo {author} {\bibfnamefont {P.~K.}\ \bibnamefont {Barkoutsos}}, \bibinfo {author} {\bibfnamefont {G.}~\bibnamefont {Nannicini}}, \bibinfo {author} {\bibfnamefont {A.}~\bibnamefont {Robert}}, \bibinfo {author} {\bibfnamefont {I.}~\bibnamefont {Tavernelli}},\ and\ \bibinfo {author} {\bibfnamefont {S.}~\bibnamefont {Woerner}},\ }\href@noop {} {\bibfield  {journal} {\bibinfo  {journal} {Quantum}\ }\textbf {\bibinfo {volume} {4}},\ \bibinfo {pages} {256} (\bibinfo {year} {2020})}\BibitemShut {NoStop}%
\bibitem [{\citenamefont {Sato}\ \emph {et~al.}(2025)\citenamefont {Sato}, \citenamefont {Gordon}, \citenamefont {Saito}, \citenamefont {Kawashima}, \citenamefont {Nikuni},\ and\ \citenamefont {Watabe}}]{sato2025two}%
  \BibitemOpen
  \bibfield  {author} {\bibinfo {author} {\bibfnamefont {R.}~\bibnamefont {Sato}}, \bibinfo {author} {\bibfnamefont {C.}~\bibnamefont {Gordon}}, \bibinfo {author} {\bibfnamefont {K.}~\bibnamefont {Saito}}, \bibinfo {author} {\bibfnamefont {H.}~\bibnamefont {Kawashima}}, \bibinfo {author} {\bibfnamefont {T.}~\bibnamefont {Nikuni}},\ and\ \bibinfo {author} {\bibfnamefont {S.}~\bibnamefont {Watabe}},\ }\href@noop {} {\bibfield  {journal} {\bibinfo  {journal} {IEEE Transactions on Quantum Engineering}\ } (\bibinfo {year} {2025})}\BibitemShut {NoStop}%
\bibitem [{\citenamefont {Liu}\ \emph {et~al.}(2025)\citenamefont {Liu}, \citenamefont {Qian}, \citenamefont {Wu}, \citenamefont {Fan}, \citenamefont {Zhang}, \citenamefont {Cai}, \citenamefont {Lu}, \citenamefont {Wang},\ and\ \citenamefont {Wang}}]{liu2025solving}%
  \BibitemOpen
  \bibfield  {author} {\bibinfo {author} {\bibfnamefont {L.}~\bibnamefont {Liu}}, \bibinfo {author} {\bibfnamefont {L.}~\bibnamefont {Qian}}, \bibinfo {author} {\bibfnamefont {X.-Y.}\ \bibnamefont {Wu}}, \bibinfo {author} {\bibfnamefont {C.-R.}\ \bibnamefont {Fan}}, \bibinfo {author} {\bibfnamefont {L.-F.}\ \bibnamefont {Zhang}}, \bibinfo {author} {\bibfnamefont {D.-B.}\ \bibnamefont {Cai}}, \bibinfo {author} {\bibfnamefont {H.-J.}\ \bibnamefont {Lu}}, \bibinfo {author} {\bibfnamefont {T.-J.}\ \bibnamefont {Wang}},\ and\ \bibinfo {author} {\bibfnamefont {C.}~\bibnamefont {Wang}},\ }\href@noop {} {\bibfield  {journal} {\bibinfo  {journal} {IEEE Transactions on Intelligent Transportation Systems}\ } (\bibinfo {year} {2025})}\BibitemShut {NoStop}%
\bibitem [{\citenamefont {Sano}\ \emph {et~al.}(2024)\citenamefont {Sano}, \citenamefont {Mitarai}, \citenamefont {Yamamoto},\ and\ \citenamefont {Ishikawa}}]{sano2024accelerating}%
  \BibitemOpen
  \bibfield  {author} {\bibinfo {author} {\bibfnamefont {Y.}~\bibnamefont {Sano}}, \bibinfo {author} {\bibfnamefont {K.}~\bibnamefont {Mitarai}}, \bibinfo {author} {\bibfnamefont {N.}~\bibnamefont {Yamamoto}},\ and\ \bibinfo {author} {\bibfnamefont {N.}~\bibnamefont {Ishikawa}},\ }\href@noop {} {\bibfield  {journal} {\bibinfo  {journal} {IEEE Transactions on Quantum Engineering}\ }\textbf {\bibinfo {volume} {5}},\ \bibinfo {pages} {1} (\bibinfo {year} {2024})}\BibitemShut {NoStop}%
\bibitem [{\citenamefont {Gilliam}\ \emph {et~al.}(2021)\citenamefont {Gilliam}, \citenamefont {Woerner},\ and\ \citenamefont {Gonciulea}}]{gilliam2021grover}%
  \BibitemOpen
  \bibfield  {author} {\bibinfo {author} {\bibfnamefont {A.}~\bibnamefont {Gilliam}}, \bibinfo {author} {\bibfnamefont {S.}~\bibnamefont {Woerner}},\ and\ \bibinfo {author} {\bibfnamefont {C.}~\bibnamefont {Gonciulea}},\ }\href@noop {} {\bibfield  {journal} {\bibinfo  {journal} {Quantum}\ }\textbf {\bibinfo {volume} {5}},\ \bibinfo {pages} {428} (\bibinfo {year} {2021})}\BibitemShut {NoStop}%
\end{thebibliography}%

\end{document}